\documentclass[aps, prd,10pt,reprint,showkeys,superscriptaddress, nofootinbib,floatfix]{revtex4-2}

\usepackage{amsmath}
\usepackage{graphicx}
\usepackage{dcolumn}
\usepackage{bm}
\usepackage{amsfonts}
\usepackage{mathrsfs}
\usepackage{amssymb}
\usepackage{slashed}
\usepackage[colorlinks=true,linkcolor=blue]{hyperref}
\usepackage[capitalize]{cleveref}
\usepackage[x11names]{xcolor}
\usepackage{bm}
\usepackage{empheq}
\usepackage{ragged2e}
\usepackage[normalem]{ulem}

\crefname{section}{Sec.}{Sec.}
\Crefname{section}{Section}{Sections}

\usepackage[super]{nth}

\def\CC{\mathcal{C}}

\usepackage{dsfont}
\usepackage{url}

\usepackage{color}
\usepackage{colortbl}

\begin{document}

\title{Shear viscosity from perturbative Quantum Chromodynamics to the hadron resonance gas at finite baryon, strangeness, and electric charge densities}

\author{Isabella Danhoni}
\email{idanhoni@theorie.ikp.physik.tu-darmstadt.de}
\affiliation{Institut für Kernphysik,
T. U. Darmstadt, Schlossgartenstraße 2,
D-64289 Darmstadt, Germany}
\author{Jordi Salinas San Mart\'in}
\affiliation{Illinois Center for Advanced Studies of the Universe, Department of Physics, University of Illinois at
Urbana-Champaign, Urbana, IL 61801, USA}
\author{Jacquelyn Noronha-Hostler}
\affiliation{Illinois Center for Advanced Studies of the Universe, Department of Physics, University of Illinois at
Urbana-Champaign, Urbana, IL 61801, USA}

\begin{abstract}
    Through model-to-data comparisons from heavy-ion collisions, it has been shown that the Quark Gluon Plasma has an extremely small shear viscosity at vanishing densities.  At large baryon densities, significantly less is known about the nature of the shear viscosity from Quantum Chromodynamics (QCD). Within heavy-ion collisions, there are three conserved charges: baryon number (B), strangeness (S), and electric charge (Q).  Here we calculate the shear viscosity in two limits using perturbative QCD and an excluded-volume hadron resonance gas at finite BSQ densities. We then develop a framework that interpolates between these two limits such that shear viscosity is possible to calculate across a wide range of finite BSQ densities. We find that the pQCD and hadron resonance gas calculations have different BSQ density dependencies such that a rather non-trivial shear viscosity appears at finite densities. 
\end{abstract}

\maketitle
\newpage

\section{Introduction}
\label{Sec:Introduction}

The Quark Gluon Plasma (QGP) produced in heavy-ion collisions at the Relativistic Heavy-Ion Collider (RHIC) or Large Hadron Collider (LHC) is a deconfined state of matter composed of quarks and gluons \cite{PHENIX:2004vcz,BRAHMS:2004adc,PHOBOS:2004zne,STAR:2005gfr}. 
Around the phase transition between quarks and gluons into hadrons, the matter is very strongly-interacting and has been shown to have an extremely small specific shear viscosity--the shear viscosity to entropy density ratio, $\eta/s$--that reaches approximately the so-called KSS limit \cite{Kovtun:2004de}, the lowest of any known substance. 
State-of-the-art Bayesian analyses \cite{Bernhard:2019bmu,JETSCAPE:2020mzn,Nijs:2020roc,Parkkila:2021tqq} using relativistic viscous hydrodynamics estimate the value of $\eta/s$ to be approximately $\eta/s\sim 0.08$--$0.2$, when considering the regime of vanishing baryon density. 
It is generally believed that there is a minimum in $\eta/s$ as a function of temperature $T$ \cite{Bernhard:2019bmu} in the strongly-interacting regime around the cross-over phase transition between quarks and gluons into hadrons and that $\eta/s$ then increases as the interactions become weaker either at high temperatures in the quark phase \cite{Arnold:2000dr,Arnold:2003zc,Ghiglieri:2018dib,Ghiglieri:2018dgf} or low temperatures in the Hadron Resonance Gas (HRG) phase  \cite{Noronha-Hostler:2008kkf,Demir:2008tr,Pal:2010es,Khvorostukhin:2010aj,Tawfik:2010mb,Alba:2015iva,Ratti:2010kj,Tiwari:2011km,NoronhaHostler:2012ug,Kadam:2014cua,Kadam:2015xsa,Kadam:2015fza,Rose:2017bjz,Kadam:2018hdo,Mohapatra:2019mcl}. 
However, extraction from data of the exact functional form of $\eta/s$ at vanishing densities presents challenges and may still yet be consistent with a constant $\eta/s$.
One should also note that the Bayesian analyses use \textit{ansätze} functional forms of $\eta/s$, not microscopic calculations and, therefore, it is not possible to easily extrapolate these results to new, yet underexplored regions of the Quantum Chromodynamics (QCD) phase diagram.

It is not yet possible to directly calculate $\eta/s$ in the strongly interacting regime from lattice QCD due to the fermion sign problem \cite{Troyer:2004ge,Ratti:2018ksb} (although quantum computers may eventually overcome this limitation \cite{Cohen:2021imf}). 
Instead, one may compute the shear viscosity in the weakly coupled regime using perturbative QCD (pQCD) techniques, as previously demonstrated at vanishing baryon densities \cite{Arnold:2000dr,Arnold:2003zc,Ghiglieri:2018dib,Ghiglieri:2018dgf} and, more recently, at finite baryon densities \cite{Danhoni:2022xmt}.  
These pQCD calculations are most applicable at high temperatures, where the QGP is expected to become weakly coupled. 
While the precise temperature at which pQCD becomes applicable remains ambiguous, at lower temperatures one can calculate the shear viscosity using the known properties of the HRG using kinetic theory. 
For instance, employing an HRG with excluded-volume corrections \cite{Gorenstein:2007mw,Noronha-Hostler:2012ycm,McLaughlin:2021dph} yields a shear viscosity that decreases with increasing temperature---a behavior consistent with the expectation of a minimum in $\eta/s$ near the cross-over phase transition. 
This trend can be understood because more degrees of freedom open up at higher temperatures, thereby increasing the entropy and reducing $\eta/s$ \cite{Noronha-Hostler:2008kkf}. 
However, HRG calculations based on the latest Particle Data Group (PDG) hadronic list (as of 2016, comprising over 700 hadrons) found the value of $\eta/s$ to be significantly larger than those suggested by Bayesian analyses \cite{McLaughlin:2021dph}. 
An alternative approach within the same temperature regime is provided by hadron transport theory \cite{Demir:2008tr,Rose:2017bjz,Hammelmann:2023fqw}; however, one would want to include the latest PDG list (see \cite{SanMartin:2023zhv}), which remains for future work. 
In the strongly coupled regime, holography has emerged as a popular approach, wherein a ``fuzzy" lower bound of $\eta/s=1/(4\pi)$ has been calculated, commonly referred to as the KSS limit \cite{Kovtun:2004de}. 
Nevertheless, there are various ways in which holographic models introduce a temperature dependence in $\eta/s$ \cite{Cremonini:2011iq} or even allow for violations of the KSS bound \cite{Kats:2007mq,Brigante:2007nu,Buchel:2008vz,Brigante:2008gz}. 
Alternatively, several non-perturbative quantum field theory techniques have been developed to calculate  $\eta/s$ as well \cite{Christiansen:2014ypa,Lowdon:2021ehf}.

Up until now, we have we have focused solely on the behavior of shear viscosity at vanishing densities, i.e., the regime applicable at top LHC energies. 
At finite baryon densities ($n_B$), $\eta/s$ is no longer the appropriate quantity of interest.
Instead, one uses $\eta T/w$, where $w=\varepsilon+p$ is the enthalpy (note that for vanishing baryon densities, $\eta T/w$ and $\eta/s$ become equivalent), with $\varepsilon$ representing the energy density and $p$ the pressure.  
The search for the QCD critical point at large baryon densities is ongoing \cite{An:2021wof,Dexheimer:2020zzs,Lovato:2022vgq,Sorensen:2023zkk}, and rich physics questions related to shear viscosity arise in this regime.
However, at large baryon densities, our understanding of the shear viscosity behavior remains limited.
Various model calculations have shown either an increase \cite{Soloveva:2020hpr} or decrease \cite{Denicol:2013nua,Grefa:2022sav} in $\eta T/w$ with increasing baryon chemical potential $\mu_B$. 
Preliminary Bayesian analyses suggest that $\eta T/w$ should increase with $\mu_B$ \cite{Auvinen:2017fjw,Shen:2023awv}, but a clear functional form has yet to emerge from the data. 
The presence of a critical point/first-order phase transition at large $\mu_B$ would also influence the behavior of $\eta T/w$. 
While the influence of any critical scaling is most likely negligible for the shear viscosity  \cite{Monnai:2016kud}, the location of the critical point could have an impact on the finite $\mu_B$ behavior, as all inflections points must converge at the critical point (i.e., the minimum of $\eta T/w$); see \cite{McLaughlin:2021dph,Grefa:2022sav} for more details. 
Additionally, a jump in $\eta T/w$ across the first-order phase transition line is anticipated after the critical point \cite{Soloveva:2020hpr,Grefa:2022sav}. 

In summary, we can expect certain behaviors of $\eta T/w$ if there is a critical point. 
However, the location and even the existence of the QCD critical point is still quite uncertain. 
Thus, we require a framework to describe $\eta T/w$ at finite $\mu_B$, which accommodates a shift on the location of the critical point or has no critical point whatsoever. 
To achieve this, we must incorporate the required features and phase transitions as described above, while allowing for flexibility in when and how these changes occur between phases. 

A further complication is that at large baryon densities other globally conserved charges play a significant role. 
For instance, due to strangeness neutrality and the existence of strange baryons and mesons in the QGP, one acquires a rather large strangeness chemical potential at finite $\mu_B$. 
In fact, $\mu_S\sim \mu_B/3$ when one enforces strangeness neutrality. 
Moreover, electric charge also plays a role since the nuclei that are stopped contain a large number of protons that carry electric charge. 
The ratio of the number of protons $Z$ to the total number of nucleons $A$ in the nuclei remains fixed globally throughout the lifetime of the QGP such that $Z/A=\text{const}$. 
For heavy nuclei used in relativistic heavy-ion collisions the values range from  $Z/A\sim 0.38$--$0.45$. 
Enforcing conservation of electric charge leads to a small, negative $\mu_Q$ that is approximately one tenth of $\mu_B$ in magnitude, i.e., $\mu_Q\sim -\mu_B/10$ \cite{Monnai:2021kgu}. 
Thus far, we have only discussed the global values of the chemical potentials, but it is also possible to have local fluctuations of conserved charges (as long as the strangeness neutrality and electric charge conservation are enforced globally); see examples \cite{Carzon:2019qja,Plumberg:2023vkw,Plumberg:2024leb}. 
In those cases, you actually require an equation of state that is 4-dimensional such that it depends on temperature, baryon ($B$) chemical potential, strangeness ($S$) chemical potential, and electric charge ($Q$) chemical potential: $\left\{T,\mu_B,\mu_S,\mu_Q\right\}$ and must also have a description for $\eta T/w\left(T,\mu_B,\mu_S,\mu_Q\right)$. 
Currently, such a description does not exist. 

The paper is organized as follows.
In Sec.~\ref{Sec:Calculation_of_shear_viscosity_at_finite_temperatures_and_densities} we first present the basic thermodynamic relations that will be used throughout this paper and introduce the procedure to calculate $\eta T/w$ in the case of multiple conserved charges in two limits: pQCD with 3 flavors and HRG with an excluded-volume. 
Then, in Sec.~\ref{Sec:Building_a_phenomenological_model_of_three_conserved_charges} we discuss the applicability of both limits around the first-order transition line and use comparisons to lattice QCD to motivate the points of switching between the HRG to an intermediate regime and from this to pQCD. 
Later, we establish the framework to connect the pQCD and HRG model calculations, wherein we use various functional forms to interpolate between these two limits.  
Additionally, we renormalize the magnitude of $\eta T/w$ for pQCD calculations at vanishing densities to match NLO calculations from GMT \cite{Ghiglieri:2018dgf} and introduce an overall renormalization based on the KSS bound, similar to what was done in previous work \cite{McLaughlin:2021dph}. 
In Sec.~\ref{Sec:Results_for_etaT_over_omega}, we present the first calculation of $\eta T/w$ for multiple conserved charges across the QCD phase diagram. 
Our results demonstrate the non-trivial effect of $\mu_B$, $\mu_S$, and $\mu_Q$ on $\eta T/w$, which opens up an avenue for the investigation of these effects on hydrodynamic calculations.

\section{Calculation of shear viscosity at finite temperatures and densities}
\label{Sec:Calculation_of_shear_viscosity_at_finite_temperatures_and_densities}

In this section, we present the shear viscosity calculations at high temperatures using perturbative QCD (pQCD) and at low temperatures using a hadron resonance gas (HRG). 

\subsection{Thermodynamics} 
\label{Sec:Thermodynamics}
\label{thermo}

We begin this section by introducing the thermodynamics of an ideal, relativistic gas of point-like particles with multiple conserved charges, which will be used for both the weakly interacting QCD phase and the hadron resonance gas phase. In \cref{sec:HRG}, we will also examine the potential impact of excluded-volume effects that account for repulsive interactions. Within the grand canonical ensemble, the ideal gas behaves as an uncorrelated system of free particles \cite{Venugopalan:1992hy}. For such a system, the partition function can be written as
\begin{equation}
    \label{eq:total_partition_function}
    \frac{T\ln Z_\mathrm{id}}{V} = \sum_i\frac{T\ln Z_\mathrm{i}}{V},
\end{equation}
where the sum runs over all particle species in the system and $Z_i$ is the partition function for each particle species $i$, and is given by
\begin{widetext}
\begin{equation}
    \frac{T\ln Z_\mathrm{i}}{V} = \frac{g_i}{6\pi^2}\int_0^\infty dp\,\frac{p^4}{\sqrt{p^2+m_i^2}}\frac{1}{\exp{\left[\left(\sqrt{p^2+m_i^2}\right)/T-\tilde{\mu}_i\right]}+\eta_i},
\end{equation}
\end{widetext}
where $\tilde{\mu}_i$ is the dimensionless effective chemical potential of particles species $i$, defined by
\begin{equation}\label{eqn:effchem}
    \tilde{\mu}_i = B_i \frac{\mu_B}{T} + S_i \frac{\mu_S}{T} + Q_i \frac{\mu_Q}{T}.
\end{equation}
Here, $g_i=2J_i+1$ is the spin degeneracy for a particle of spin $J_i$, $m_i$ is its mass, and $\eta_i=+1$ for fermions and $\eta_i=-1$ for bosons.
The quantities $B_i$, $S_i$, and $Q_i$ are the baryon number, strangeness, and electric charge of each particle, while $\mu_B$, $\mu_S$, and $\mu_Q$ are the corresponding chemical potentials of each conserved charge.

In principle, we can approximate the quark-gluon plasma with a free gas composed of gluons and quarks (and their corresponding anti-quarks) of $u$, $d$, and $s$ flavors  that, through their respective quantum numbers, lead to conservation of the $BSQ$ charges.  
In this system, \cref{eq:total_partition_function} reduces to
\begin{widetext}
\begin{align}
    \frac{T\ln Z_\mathrm{QGP}}{V}&= g_b\frac{\pi^2}{90}T^4 + \sum_{f=\{u,d,s\}} \frac{g_f}{6\pi^2}\Bigg(\frac{7\pi^4}{60} + \frac{\tilde{\mu}_f^2\pi^2}{2} + \frac{\tilde{\mu}_f^4}{4}\Bigg) T^4.
    \label{partfunc}
\end{align}    
\end{widetext}
in the limit of high temperature and chemical potentials.
Here, $g_b=2(N_c^2-1)=16$ accounts for all gluonic degrees of freedom, since gluons come in 8 possible color combinations and two polarizations.
On the other hand, the factor $g_f=2N_c=6$ is the degeneracy factor for 2 helicities and 3 quark colors. 
Lastly, the sum is over the available quark flavors. 
The contribution from anti-quarks is already included in the partition function and does not require additional counting. 

Generically, from the partition function one can calculate the thermodynamic properties of the system such as the pressure $p$, the energy density $\varepsilon$, the net number density of the $i$-th species $n_i$, and entropy density $s$ from
\begin{subequations}
    \label{eq:thermodynamics_from_partition_function}
    \begin{align}
    p(T,\mu_B,\mu_S,\mu_Q) &= \frac{T\ln Z}{V},\\
    \varepsilon(T,\mu_B,\mu_S,\mu_Q) &=\frac{T^2}{V}\frac{\partial \ln Z}{\partial T},\\
    n_i(T,\mu_B,\mu_S,\mu_Q) &= \frac{1}{V}\frac{\partial \ln Z}{\partial \tilde{\mu}_i},\\
    s(T,\mu_B,\mu_S,\mu_Q) &= \frac{1}{V}\frac{\partial T\ln Z}{\partial T}.
    \end{align}
\end{subequations}
In addition, we calculate
\begin{align}
    \label{eq:derived_thermodynamics}
    \frac{\rho_X (T,\mu_B,\mu_S,\mu_Q)}{T^3}&=\frac{\partial p(T,\mu_B,\mu_S,\mu_Q)/T^4}{\partial (\mu_X/T)},
\end{align}
where $\rho_X$ is the net density associated with a specific conserved charge $X =\{B, S, Q\}$.
Finally, all thermodynamic quantities in Eqs.~\eqref{eq:thermodynamics_from_partition_function} and \cref{eq:derived_thermodynamics} can be connected through the Gibbs-Duhem relation,
\begin{equation}
    \varepsilon+p= sT + \sum_{X=\{B,S,Q\}}\mu_X\rho_X.
\end{equation}

\subsection{Perturbative QCD}
\label{pQCD}

The transport coefficients of a weakly coupled QCD medium can be derived using a kinetic theory framework. 
The first correct calculation was performed by Arnold, Moore, and Yaffe \cite{Arnold:2000dr,Arnold:2003zc}, and later expanded by Danhoni and Moore \cite{Danhoni:2022xmt} to the high-density region for massless quarks. 
In the present study, we apply the results found for shear viscosity as a function of $\mu_B$ and adopt the same formalism to obtain $\eta$ at finite $\mu_S$ and $\mu_Q$. 
Below, we briefly review the important elements of this calculation and discuss the specific modifications relevant to our case. 

As noted earlier, these calculations employ a kinetic theory description for the hot, dense QCD matter in a weakly coupled regime. 
In this picture, shear viscosity is a property relevant for systems with space-nonuniform flow velocity, and the local equilibrium distribution for fermions and bosons is given by the Fermi-Dirac and Bose-Einstein distributions, respectively,
\begin{equation}
\label{f0}
    f_0^{q/\overline{q}} = \frac{1}{\exp\left(\beta p - \tilde{\mu_i}\right) + 1}, \qquad f_0^g = \frac{1}{\exp\left(\beta p\right) - 1}.
\end{equation}
\noindent
Here $\beta^{-1} = T$ is the local temperature. 
In order to describe a generic state that is not necessarily in equilibrium, we then require a distribution that includes both the equilibrium $f_0^a$ and out-of-equilibrium contributions $f_1^a$. 
This system has a non-equilibrium distribution that is given by
\begin{equation}
\label{f1}
    f^a(\vec k,\vec x) = f_0^a(\vec k, \vec x) + f_0^a(1 \pm f_0^a)\, f_1^a(\vec k,\vec x) \,,
\end{equation}
in which $a$ represents the species considered (a gluon, quark, or anti-quark) and the dynamics of the non-equilibrium distribution $f^a$ are determined from a Boltzmann-type equation, where time derivatives and external forces are not relevant. 
Thus, we can simply write it as
\begin{equation}
    \vec{v}_k \cdot \frac{\partial}{\partial \vec{x}}  f^a(\vec{k},\vec{x},t) = - \CC^a[f].
    \label{boltz}
\end{equation}
Here, we neglect thermal and Lagrangian masses, making the velocity vector the unit vector in the direction of the momentum, $\vec{v}_p = \hat{p} \equiv \vec p / p$. We note that the massless-quark approximation is justified in this calculation for strange quarks since the relevant energy scale here is given by the momentum of the particles, where $m_s\approx T \ll p $ (around $3T$). Therefore, a finite strange quark mass would lead to small corrections that can be neglected at leading-log order calculations. We also only consider first-order contributions to the collision operator, which takes the form

\begin{figure}[htbp]
    \centering
    \includegraphics[width=0.9\columnwidth]{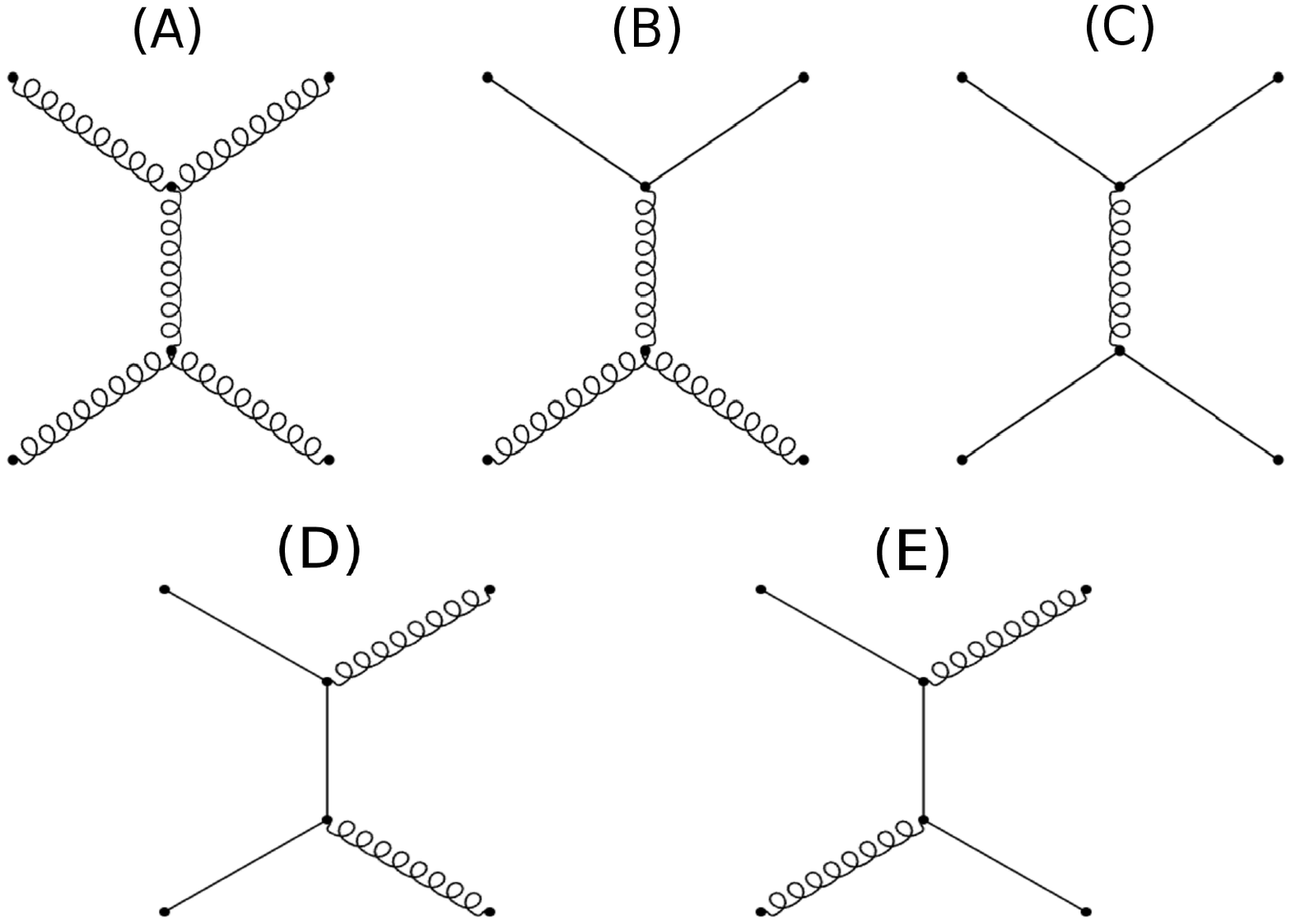}
    \caption{Feynman diagrams of the scattering contributing at leading logarithmic order \cite{Danhoni:2022xmt}.}
    \label{diagrams}
\end{figure}

\begin{widetext}
\begin{align}
    \nonumber
    \CC^a[f](\vec{p}) &=\frac{1}{2} \sum_{bcd} \int_{\vec{k},\vec{p},\vec{k'}} 
    \frac{|\mathcal{M}_{abcd}(P,K,P',K')|^2}{2p^0\, 2k^0\, 2p'{}^0\, 2k'{}^0} (2\pi)^4\, \delta^4(P+K-P'-K') \\
    &\qquad \times\Big\{ f^a(\vec{p})f^b(\vec{k})[1\pm f^c(\vec{p'})][1\pm f^d(\vec{k'})] -  f^c(\vec{p'})f^d(\vec{k'})[1\pm f^a(\vec{p})][1\pm f^b(\vec{k})]\Big\},
    \label{collop22}
\end{align}    
\end{widetext}
\noindent
where $\mathcal{M}$ represents the scattering matrix of the processes of interest, calculated using perturbative QCD; the relevant diagrams for shear viscosity at leading logarithmic order are shown in \cref{diagrams}. The sum in \cref{collop22} is over all scattering processes in the QGP of type $a + b \to c +d $. A detailed description of how to compute the contribution from each diagram and how to use the variational method to obtain shear viscosity can be found in \cite{Danhoni:2022xmt}. It is straight forward to extend this procedure for the case of multiple chemical potentials.
As a consequence of having multiple charges, different flavors of quarks and anti-quarks must have different relaxation functions, i.e.,
\begin{align}
\chi^u(k)&=\sum_{m=1}^N a_{m+N} \phi^{(m)}(k),\nonumber\\
\chi^d(k)&=\sum_{m=1}^N a_{m+2N} \phi^{(m)}(k),\nonumber\\
\chi^s(k)&=\sum_{m=1}^N a_{m+3N} \phi^{(m)}(k),
\end{align}
for quarks,
\begin{align}
\chi^{\overline{u}}(k)=\sum_{m=1}^N a_{m+4N} \phi^{(m)}(k),\nonumber \\
\chi^{\overline{d}}(k)=\sum_{m=1}^N a_{m+5N} \phi^{(m)}(k),\nonumber \\
\chi^{\overline{s}}(k)=\sum_{m=1}^N a_{m+6N} \phi^{(m)}(k),
\end{align}
for anti-quarks, and
\begin{equation}
\chi^g(k)=\sum_{m=1}^N a_m \phi^{(m)}(k), 
\end{equation}
for gluons, where $a_m$ is a variational parameter and $N$ is the size of the basis set one uses to solve the Boltzmann equation. 
In this paper, we use the same basis set as in \cite{Danhoni:2022xmt}: 
\begin{equation}
    \phi^{(m)}(k)=\frac{k(k/T)^m}{(1+k/T)^{N-1}}, \qquad m=1,...,N.
\end{equation}
This change implies that the contribution from each quark in all diagrams has to be counted separately; therefore, the scattering matrix will evolve differently as a function of each chemical potential. As discussed previously, at finite densities, the shear viscosity is normalized by $T/w$ such that we require knowledge of the enthalpy, i.e., $w=\varepsilon+p$.

\begin{figure}
    \centering
    \includegraphics[width=0.99\columnwidth]{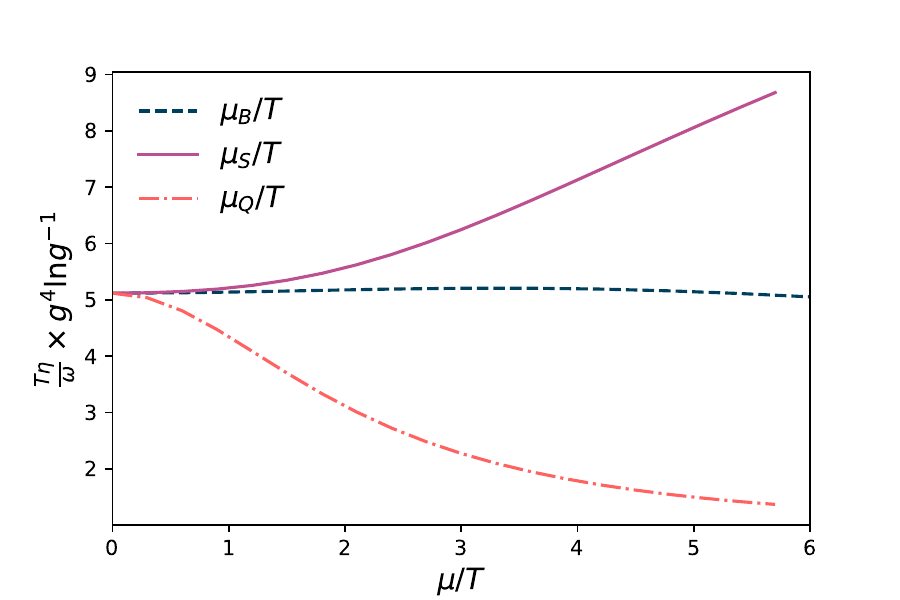}
    \caption{(color online) $\eta T/w$ at finite charge chemical potentials as a function of $\mu_B/T$ (dashed blue), $\mu_S/T$ (solid purple), and $\mu_Q/T$ (dot-dashed red) from pQCD. Every contribution is calculated while fixing the other chemical potentials to zero.
    }
    \label{fig:mu_s_shear}
\end{figure}

\Cref{fig:mu_s_shear} shows an $\eta T/w$ curve for each of $\mu_B/T$, $\mu_S/T$, and $\mu_Q/T$ while leaving the other chemical potentials set to zero. 
The curves exhibit completely different behaviors, suggesting that shear viscosity is strongly dependent on the chemical potential of the system. 
We find that $\eta T/w$ increases with increasing $\mu_S$ but decreases with increasing $\mu_Q$. 
Previous work on finite $\mu_B$ only found that  $\eta T/w$  slightly increased before generally decreasing at $\mu_B/T\gtrsim 1$ \cite{Danhoni:2022xmt}. 
Note that the values of $\mu_S$ and $\mu_Q$ here are chosen to be positive, but the results should have reflection symmetry in $\mu$ when only one chemical potential is finite.
Thus, all three chemical potentials on their own lead to strikingly different behaviors.
These differences can be explained by the fact that $\rho_Q$ grows slower as a function of $\mu_Q$ than what would happen for $\rho_B$ and $\rho_S$, and this affects the enthalpy, which will change significantly less than for the same value of $\mu_B$ or $\mu_S$. 

\begin{figure*}
\begin{minipage}{.5\textwidth}
    \centering
    \includegraphics[width=1.0\linewidth]{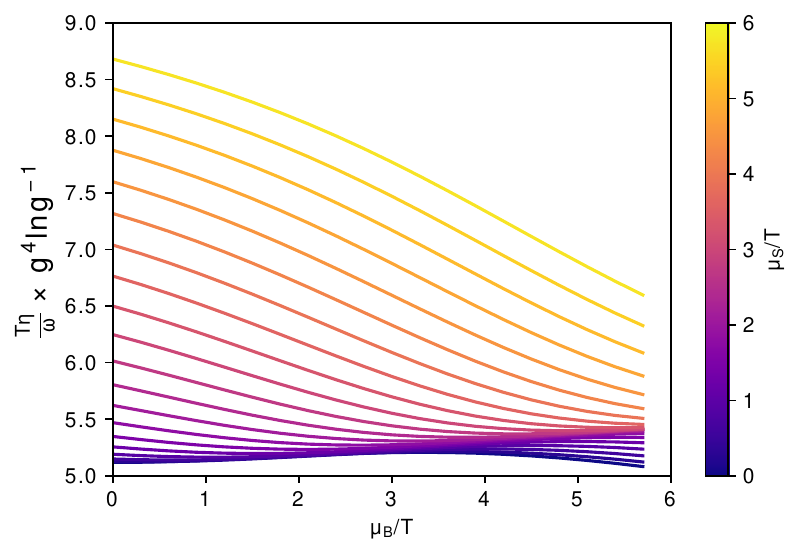}
   
    \end{minipage}%
\begin{minipage}{.5\textwidth}
    \centering
    \includegraphics[width=1.0\linewidth]{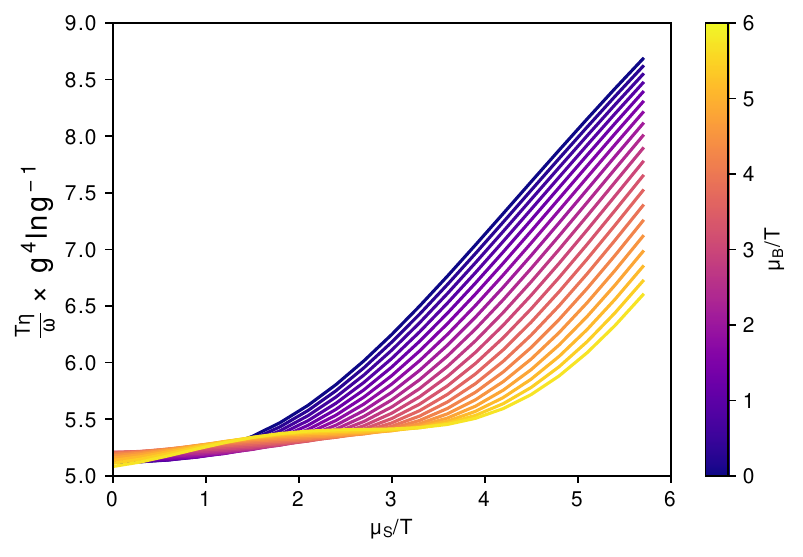}
    \end{minipage}
    \caption{(color online) Behavior of $\eta T/w$ as a function of (left) $\mu_B$ for varying $\mu_S$ and (right) $\mu_S$ for varying $\mu_B$ for the pQCD calculations performed here. In both cases, $\mu_Q=0$.
    }
     \label{shear_mus_mub}
\end{figure*}

We next hold $\mu_Q=0$ fixed and vary $\mu_B$ and $\mu_S$ to understand the interplay between chemical potentials.
In \cref{shear_mus_mub} (left), we plot $\mu_B/T$ on the $x$-axis while varying $\mu_S/T$ using different colors (shown in the color bar to the right of the plot). Also, in \cref{shear_mus_mub} (right)  we vary $\mu_S/T$ on the $x$-axis while varying $\mu_B/T$ using different colors. 
When varying $\mu_B/T$ on the $x$-axis, we find that the overall $\eta T/w$ can have a variety of qualitative behaviors across $\eta T/w$. For instance, $\eta T/w$ may increase then decrease, decrease then increase or decrease only ––all depending on what is the corresponding value of $\mu_S/T$. In general, though, increasing the value of $\mu_S/T$ leads to an overall increase in the magnitude of $\eta T/w$.
On the other hand, in the case of varying $\mu_S/T$ on the $x$-axis, the qualitative behavior of $\eta T/w$ is not as strongly affected by different values of $\mu_B/T$, showing a monotonic increase in $\eta T/w$ for the chemical potential ranges considered. 
However, we do find that a smaller $\mu_B/T$ ratio leads to a larger overall increase in $\eta T/w$ at large $\mu_S/T$, whereas a larger $\mu_B/T$ ratio tends to dampen the steep rise in $\eta T/w$ at large $\mu_S/T$.
Generally, we expect that systems with large $\mu_B$ also have a reasonably large $\mu_S$ such that we anticipate a modification of $\eta T/w$ due to the effects of strangeness. 

\begin{figure}[hbtp]
    \centering
    \includegraphics[width=0.99\columnwidth]{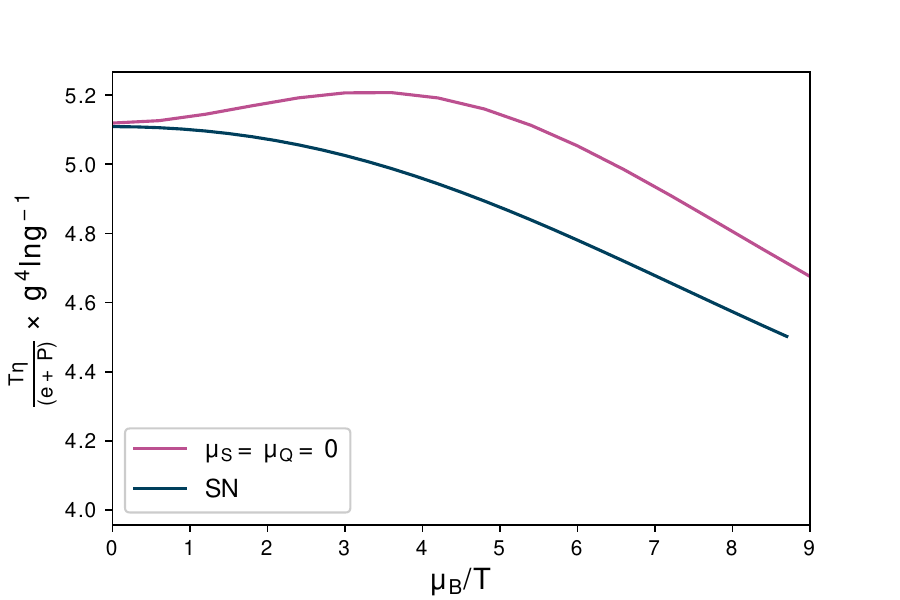}
    \caption{(color online) Shear viscosity as a function of $\mu_S(T,\mu_B)$ and $\mu_Q(T,\mu_B)$ for (black) the strangeness neutral case that has the conditions shown in \cref{eqn:electriccharge,eqn:str_neutral} compared to (pink) the case of $\mu_S=\mu_Q=0$. 
    }
    \label{3dfig}
\end{figure}

To get a better understanding of what is probed in  heavy-ion collision experiments,  we explore the "strangeness neutral" (SN) case where the two following conditions are enforced:
\begin{align}
    \langle \rho_Q\rangle &= \frac{Z}{A} \langle \rho_B\rangle, \label{eqn:electriccharge} \\
    \langle \rho_S\rangle &= 0,\label{eqn:str_neutral}
\end{align}
\noindent
where the first constraint comes from the conservation of electric charge and the second comes from strangeness neutrality. In principle, we have 4 degrees of freedom $\left\{T,\mu_B,\mu_S,\mu_Q\right\}$. Still, for the strangeness neutral trajectories through the QCD phase diagram, then  $\mu_S(T,\mu_B)$ and $\mu_Q(T,\mu_B)$ are constrained to be single values for a specific temperature and baryon chemical potential.  
In \cref{3dfig}, we compare $\eta T/w$ for the case of only finite $\mu_B$ (pink) compared to the strangeness neutral trajectory (black) where we set $Z/A=0.4$.
For finite $\mu_B$, there is a non-monotonic behavior in $\eta T/w$.
However, once we enforce the strangeness neutrality conditions, the non-monotonicity disappears, and $\eta T/w$ only decreases with increases $\mu_B$.
The effect of the strangeness neutral case is consistent with \cref{shear_mus_mub} (left) wherein if we have both finite $\mu_B$ and a sufficiently large $\mu_S$, we find that the effect of $\mu_S$ can lead to a monotonically decreasing $\eta T/w$.

\subsection{Hadron Resonance Gas Model with an Excluded-Volume}
\label{sec:HRG}
The statistical approach to describe experimental relativistic heavy-ion collision data, realized in the form of the ideal HRG model, has been used to successfully reproduce hadron yields of various species and their corresponding ratios, as well as extracting thermodynamic information at freeze-out of the strongly interacting medium generated in heavy-ion reactions, e.g., temperature $T$, chemical potential $\mu_B$ and system volume $V$. Furthermore, extensions to the HRG model such as the excluded volume hadron resonance gas (EV-HRG) and van der Waals (vdW) equation of state introduce repulsive and attractive interactions between hadrons, which have been shown to improve the effectiveness of models to describe lattice QCD data like partial pressures and susceptibilities \cite{Vovchenko:2014pka,Vovchenko:2016rkn,Alba:2016hwx,Satarov:2016peb}. At the same time, the HRG model has proven useful in explaining the liquid-gas phase transition at low temperatures in the QCD phase diagram \cite{Vovchenko:2016rkn,Lysenko:2025fex}. The simplicity of the model lies in the reduced number of free parameters. Despite this, the HRG model is able to explore strongly-interacting confined matter, the low-temperature and high-baryon density regime of the QCD phase diagram.

At low temperatures in heavy-ion collisions, the QGP freezes-out into a gas of interacting hadrons.  For years groups have studied the thermalization of the HRG.  Some examples across the years include \cite{Rapp:2000gy,Greiner:2000tu,Greiner:2004vm,Noronha-Hostler:2007fzh,Noronha-Hostler:2009wof,Noronha-Hostler:2014usa,Noronha-Hostler:2014aia,Beitel:2016ghw,Gallmeister:2017ths,Mazeliauskas:2018irt,Kurkela:2018xxd,Mazeliauskas:2019ifr}.
The general consensus is that on very short timescales chemical equilibrium is reached within the HRG due to both multi-particle interactions and the large number of heavy resonances (beyond 2 body interactions). 

In this regime, we can use the hadron resonance gas model, where we include repulsive interactions through an excluded-volume (EV-HRG) such that the hard-core volume of each hadron $i$ is 
\begin{equation}
    v_i=4\cdot\frac{4}{3}\pi r_i^3,
\end{equation}
where $r_i$ is the radius of an individual hadron. The factor of $4$ is because the excluded-volume for a sphere is 8 times its volume, but this is shared across two interacting particles, hence a factor of $4$.
Here, we make the simple assumption that all hadrons have the same volume, i.e., $v=v_i$ for all $i$. While it is possible to relax this assumption (see, for example, Ref. \cite{Albright:2014gva}), the scaling of hadron eigenvolumes is not yet well understood \cite{Alba:2016hwx}, the role of baryon, strange and electric charges is undetermined \cite{Bugaev:2012wp,Vovchenko:2015cbk,Vovchenko:2016rkn,Satarov:2016peb,Vovchenko:2017xad}, and the lack of lattice QCD and experimental data renders such approach too unconstrained.

Before including the excluded-volume, we begin with the ideal hadron resonance gas model where the equation of state is described using a gas of free hadrons (see \cref{thermo}). 
With this model, one can obtain thermodynamic quantities and transport coefficients assuming that hadrons are point-like particles. The total density of particles for species $i$ can be calculated from the partition function such that
\begin{widetext}
\begin{equation}
    \frac{n_i^\mathrm{id}(T,\mu_B,\mu_S,\mu_Q)}{T^3} = \frac{g_i}{2\pi^2}\int_0^\infty dk \;k^2 \left[\exp{\left(\frac{\sqrt{k^2+m_i^2}}{T}-\tilde{\mu}_i\right)}+(-1)^{B_i-1}\right]^{-1},
    \label{eqn:n_hrg}
\end{equation}
\end{widetext}
where $B_i$ is the baryon number of the particle, $k$ is the momentum, $m_i$ is the mass of the particle, and the degeneracy factor $g_i=2J_i+1$ comes from spin $J_i$.
The effective chemical potential for species $i$ is defined in the same way as above in \cref{eqn:effchem}.

Next, we include repulsive interactions through an excluded-volume.
The excluded volume has been used regularly within the field of heavy-ion collisions, see examples of other works \cite{Rischke:1991ke,Yen:1997rv,Gorenstein:1999ce,Vovchenko:2016ebv} and in \cite{Noronha-Hostler:2012ycm} where the Lambert function was used.
The excluded-volume pressure is given by the self-consistent equation \cite{Rischke:1991ke,Yen:1997rv,Gorenstein:1999ce,Vovchenko:2016ebv} 
\begin{align}
\nonumber
    \frac{p^\mathrm{ex}(T,\mu_B,\mu_S,\mu_Q)}{T} &= n^\mathrm{id}(T,\mu_B,\mu_S,\mu_Q)\\
    &\times\exp{\left(\frac{-v\, p^\mathrm{ex}(T,\mu_B,\mu_S,\mu_Q)}{T}\right)},
\end{align}
where $p^\mathrm{ex}$ is our excluded-volume pressure. If all hadrons have the same $v$, this self-consistent equation can be solved analytically using the Lambert $W$
function,
\begin{equation}\label{eqn:p_ex}
    p^\mathrm{ex}(T,\mu_B,\mu_S,\mu_Q) = \frac{T}{v} W(v\; n^\mathrm{id}(T,\mu_B,\mu_S,\mu_Q)).
\end{equation}
Here, $n^\mathrm{id}(T,\mu_B,\mu_S,\mu_Q)$ is the number density defined in \cref{eqn:n_hrg}. The remaining thermodynamic quantities can be derived using the relationships in \cref{thermo} (we specifically require the enthalpy $w$ to normalize the shear viscosity). 
We consider that all particles have the same volume but different mass $m_i$ and width $\Gamma_i$ \cite{Vovchenko:2018fmh}, such that in the limit $v\rightarrow 0$, one should recover the ideal gas description \cite{Noronha-Hostler:2012ycm,McLaughlin:2021dph}. 
The calculations for the thermodynamic variables were performed using the \textsc{Thermal-FIST} package \cite{Vovchenko:2019pjl} with the PDG2021+ particle list \cite{SanMartin:2023zhv}.

Now, we can discuss the calculations of the shear viscosity.
In kinetic theory, the shear viscosity is proportional to \cite{Gorenstein:2007mw,Noronha-Hostler:2012ycm}
\begin{equation}\label{eqn:eta_kinetictheory}
    \eta \propto n_\mathrm{tot}\sum_i \lambda_i \langle |k|\rangle_i,
\end{equation}
where $n_\mathrm{tot}$ is the total particle number density defined as
\begin{equation}
    n_\mathrm{tot}=\sum_i n_i,
\end{equation}
where $\lambda_i$ is the mean free path such that $\lambda \propto 1/(n_ir^2)$ and $\langle |k|\rangle_i$ is the average thermal momentum of particle $i$.  
Note that the total particle number density $n_\mathrm{tot}$ here is not the same as the net-baryon density $\rho_B$.  
Rather, to calculate the baryon density, we would need to include a weight for each particle species according to its baryon number (e.g., $B_p=1$ and $B_{\bar{p}}=-1$ for (anti)protons) such that at vanishing chemical potentials, $\rho_B=0$.  
Since we know that $\eta$ does not vanish at vanishing $\tilde{\mu}=0$, instead what enters is the sum of all particles and anti-particles without a weight with the baryon density (such that mesons are also included).
Thus, the total particle density elucidates the amount total matter plus anti-matter in the system, not the net amount of matter.

In the hadron resonance gas model, we can describe the average thermal momentum as:
\begin{equation}\label{eqn:thermal_mom_initial}
    \langle |k|\rangle_i = \frac{\int_0^\infty k^3 \exp\left[-\sqrt{k^2+m_i^2}/T+\tilde{\mu}_i\right] dk}{\int_0^\infty k^2 \exp\left[-\sqrt{k^2+m_i^2}/T+\tilde{\mu}_i\right] dk}.
\end{equation}
However, because the $\tilde{\mu}_i$ term is independent of the momentum, we can pull it out of the integral such that it cancels out and \cref{eqn:thermal_mom_initial} simplifies to
\begin{equation}\label{eqn:thermal_mom_simplified}
    \langle |k|\rangle_i = \frac{\int_0^\infty k^3 \exp\left[-\sqrt{k^2+m_i^2}/T\right] dk}{\int_0^\infty k^2 \exp\left[-\sqrt{k^2+m_i^2}/T\right] dk}
\end{equation}
even at finite chemical potentials. 
Substituting this into \cref{eqn:eta_kinetictheory}, we obtain
\begin{equation}
    \eta \propto \frac{1}{r^2}\sum_i \frac{n_i}{n_{tot}}\frac{\int_0^\infty k^3 \exp\left[-\sqrt{k^2+m_i^2}/T\right] dk}{\int_0^\infty k^2 \exp\left[-\sqrt{k^2+m_i^2}/T\right] dk}
\end{equation}
if we continue to assume a single hard-core radius for all hadrons (otherwise, $r$ would remain inside the summation). 
Next, we focus specifically on the particle number density $n_i$. 
For an ideal gas, the particle number density is shown in Eq.\ (\ref{eqn:n_hrg}) for species $i$.  
When considering excluded-volume effects, we can then write
\begin{equation}\label{eqn:excluded_vol_numb}
    n_i^\mathrm{ex}=\frac{e^{-v p^\mathrm{ex}/T}n_i^\mathrm{id}}{1+ve^{-v p^\mathrm{ex}/T}n_\mathrm{tot}^\mathrm{id}},
\end{equation}
where the total excluded-volume particle number density is 
\begin{align}
   n_\mathrm{tot}^\mathrm{ex}&= \sum_i n_i^\mathrm{ex}\nonumber\\
   &=\sum_i \frac{e^{-v p^\mathrm{ex}/T}n_i^\mathrm{id}}{1+ve^{-v p^\mathrm{ex}/T}n_\mathrm{tot}^\mathrm{id}} \nonumber\\
\end{align}
and assuming still that the excluded-volume $v$ is the same for all particle species, then we pull out the $n_i^\mathrm{id}$ term from the summation such that
\begin{equation}
       n_\mathrm{tot}^\mathrm{ex} = n_\mathrm{tot}^\mathrm{id}\frac{e^{-v p^\mathrm{ex}/T}}{1+ve^{-v p^\mathrm{ex}/T}n_\mathrm{tot}^\mathrm{id}}.
\end{equation}
Therefore, the ratios 
\begin{equation}
    \frac{n_i^\mathrm{ex}}{n_\mathrm{tot}^\mathrm{ex}}= \frac{n_i^\mathrm{id}}{n_\mathrm{tot}^\mathrm{id}}
\end{equation}
are equal when the volumes are the same.  If one relaxes this assumption, one would need to solve \cref{eqn:excluded_vol_numb} with a volume dependency on particle species $i$.

Now, we are able to finally write the shear viscosity at finite $\tilde{\mu}$ using an excluded-volume description for a hadron resonance gas
\begin{equation}
    \eta^\mathrm{HRG} = \frac{5}{64\sqrt{8}}\frac{1}{r^2}\frac{1}{n_\mathrm{tot}^\mathrm{id}}\sum_i n_i^\mathrm{id}\frac{\int_0^\infty  k^3 \exp{\left(\frac{-\sqrt{k^2 + m_i^2}}{T}\right)} dk}{\int_0^\infty  k^2 \exp{\left(\frac{-\sqrt{k^2 + m_i^2}}{T}\right)} dk},
    \label{shear_hrg}
\end{equation}
where the prefactors were derived in \cite{LIFSHITZ19811,Gorenstein:2007mw}.  
The normalization of $\eta^\mathrm{HRG}$ at finite chemical potentials, usually performed with the entropy density at vanishing chemical potentials, requires an excluded-volume calculation for the pressure $p^\mathrm{ex}$, as shown already in \cref{eqn:p_ex}, and the energy density $\varepsilon^\mathrm{ex}$, 
\begin{equation}\label{eqn:e_ex}
    \varepsilon^\mathrm{ex}\left(T,\mu\right)=\frac{\varepsilon^\mathrm{id}\left(T,\mu\right)}{\exp{\left[vp^\mathrm{ex}\left(T,\mu\right)/T\right]}+v n_\mathrm{tot}^\mathrm{id}\left(T,\mu\right) },
\end{equation}
where $\mu$ can include all three chemical potentials; see \cite{Vovchenko:2014pka} for more details. 

Finally, using \cref{eqn:p_ex,eqn:e_ex} we compute
\begin{equation}
    \frac{\eta^\mathrm{HRG} \;T}{\varepsilon^\mathrm{ex}+p^\mathrm{ex}}= \frac{\eta^\mathrm{HRG} \;T}{w^\mathrm{ex}},
\end{equation}
or, in the limit of $\mu\rightarrow 0$ one can simply calculate  $\eta^\mathrm{HRG}/ s^\mathrm{ex}$ where the entropy density is $s^\mathrm{ex}=\partial p^\mathrm{ex}/\partial T|_{\mu}$.  
From this point onwards, we will simply discuss $\eta/s$ or $\eta T/w$ (at finite $\tilde{\mu}$) and drop the index indicating that it has been derived using the excluded-volume formalism. 

The radius for the excluded-volume has been tuned to reproduce the lattice QCD entropy (see \cite{McLaughlin:2021dph} for the procedure using the PDG2016+ particle list).  
Here, we use the PDG2021+ particle list, which includes somewhat more particles. 
However, these are predominately heavier particles such that their influence does not play a substantial role in calculating the entropy (rather, their influence shows up more in the partial pressures \cite{SanMartin:2023zhv}). 
Thus, the previously used range in radii from $r=0.1$ to $0.25$ fm in \cite{McLaughlin:2021dph} works well in our study here.

\begin{figure}[htbp]
    \centering
    \includegraphics[width=0.99\columnwidth]{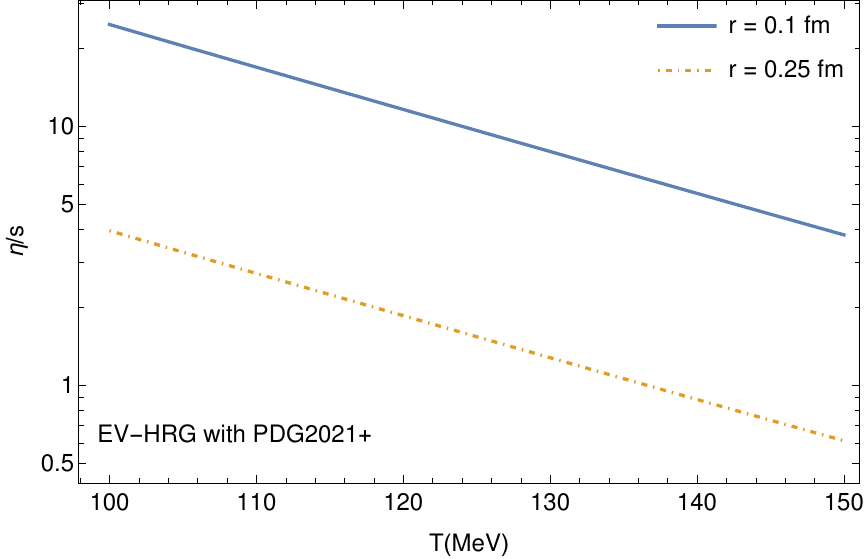}
    \caption{(color online) $\eta/s$ from the EV-HRG as a function of the temperature before any rescaling is used. The excluded-volume result for the PDG2021+ list for $r = 0.1$ fm (solid blue) is compared to $r = 0.25$ fm (dash-dotted yellow), both calculated for $\tilde{\mu}=0$.
    }
    \label{ratio}
\end{figure}

\Cref{ratio} shows a comparison of $\eta/s$ calculated directly from the HRG between the two choices of radius size, $r=0.1$ fm and $r=0.25$ fm in the temperature range relevant for the HRG. 
The plot makes it clear that shear viscosity is highly dependent on the choice of $r$ wherein the larger $r$ leads to a smaller $\eta/s$.
If we compare \cref{ratio} to Fig.\ 3 in Ref.\ \cite{McLaughlin:2021dph}, we can indeed see that the PDG2021+ lowers $\eta/s$.
For instance, for the 2016+ list in \cite{McLaughlin:2021dph}, $\eta/s\sim1$ at $T=160$ MeV whereas we find $\eta/s\sim0.4$ at $T=160$ MeV for $r=0.25$ fm. 
Thus, adding new hadronic states lowers $\eta/s$ in our framework, which is consistent with previous results \cite{Noronha-Hostler:2008kkf}. 
We also note that $r=0.1$ fm produces an $\eta/s$ that is extremely large compared to what is anticipated from phenomenology. 
However, $r=0.25$ fm is significantly closer to what we anticipate from extractions of $\eta/s$ from experimental data.

\section{Building a phenomenological model of three conserved charges}
\label{Sec:Building_a_phenomenological_model_of_three_conserved_charges}

In the sections previous to Sec. III, we explained how shear viscosity can be computed for different regimes using well-established theories, such as pQCD calculations for the high-temperature, deconfined region, and the EV-HRG model in the confined case. However, there is no theoretical guidance on how these two calculations--applicable in different regions of the phase diagram--should be connected, especially with the addition of multiple conserved charges. Thus, we opt for a phenomenological model that makes use of established frameworks, i.e., perturbative and hadron gas calculations, to provide an estimate of the shear viscosity near the deconfinement transition line that can be tested in hydrodynamic simulations. We use a simple approach wherein we match the two regions (one at high $T$ and one at low $T$) with a polynomial interpolation between the pQCD and the EV-HRG models. 
The motivation is to provide a 4D table (or tables) that can be later tested within relativistic viscous hydrodynamics calculations with BSQ-conserved charges. 
We recognize this can be updated and refined later in future work. In this paper, we intend to take a first look at $\eta T/w$ across the QCD phase diagram with finite charge densities.

\subsection{Switching between models}
\label{Subsec:Switching_between_models}

The region close to the first-order transition line cannot be described by either perturbative QCD or an HRG model; thus, we prefer to use an interpolation function to connect the two models. 
However, there is some freedom on where this interpolation should be applied. 
To deal with that, we first consider the chiral phase transition line with just finite $\mu_B$ as it is usually defined  using a Taylor expansion over the switching temperature:
\begin{equation}\label{eqn:Bchiral}
    \frac{T_\mathrm{sw}(\mu_B)}{T_{\mathrm{sw},0}} = \left[1-\kappa_2\left(\frac{\mu_B}{T_{\mathrm{sw},0}}\right)^2-\kappa_4\left(\frac{\mu_B}{T_{\mathrm{sw},0}}\right)^4\right],
\end{equation}
where $T_\mathrm{sw}$ is the chiral transition temperature, which varies with $\mu_B$, $T_{\mathrm{sw},0}$ is the chiral transition temperature at $\mu_B=0$, and $\kappa_2$ and $\kappa_4$ are dimensional parameters that determine the curvature of the transition line in the $(\mu_B, T_\mathrm{sw})$ plane and whose central values are  $\kappa_2\sim  0.015$--$0.016$, $\kappa_4\sim  0.0003$--$0.001$ as taken from lattice QCD from the Wuppertal-Budapest Collaboration \cite{Borsanyi:2011bn,Bellwied:2015rza,Borsanyi:2020fev} and the HotQCD collaboration \cite{HotQCD:2018pds}.  
The transition line drawn by \cref{eqn:Bchiral} marks the hypothetical limit between the confined and deconfined phases. 

In principle, the extension of \cref{eqn:Bchiral} to the case of three conserved charges is easily obtained, i.e., 
\begin{widetext}
\begin{equation}\label{eqn:BSQchiral}
    \frac{T_\mathrm{sw}(\mu_B,\mu_S,\mu_Q)}{ T_{\mathrm{sw},0}} =\left\{1-\sum_{i,j,k}\sum_{X,Y,Z=\{B,S,Q\}}\left[\kappa_{i+j=2}^{XY}\left(\frac{\mu_X^i\mu_Y^j}{T_{\mathrm{sw},0}^2}\right)+\kappa_{i+j+k=4}^{XYZ}\left(\frac{\mu_X^i\mu_Y^j\mu_Z^k}{T_{\mathrm{sw},0}^2}\right)\right]\right\},
\end{equation}

\noindent
where mixed terms for BSQ  may appear. 
However, current lattice QCD results \cite{HotQCD:2018pds} only include the diagonal terms for BSQ. 
Thus, Eq.\ (\ref{eqn:BSQchiral}) then simplifies to
\begin{equation}\label{eqn:BSQchiral_diag}
    \frac{T_\mathrm{sw}(\mu_B,\mu_S,\mu_Q)}{ T_{\mathrm{sw},0}} = \left\{1-\sum_{X=\{B,S,Q\}}\left[\kappa_{2}^{X}\left(\frac{\mu_X}{T_{\mathrm{sw},0}}\right)^2+\kappa_{4}^{X}\left(\frac{\mu_X}{T_{\mathrm{sw},0}}\right)^4\right]+2\kappa_{BS}\frac{\mu_B \mu_S}{T^2}\right\},
\end{equation}
\end{widetext}
\noindent
which is the formula that we will use in this work to describe the behavior of the switching temperatures across the phase diagram.
In Table \ref{tab:kappas} we list the $\kappa_n^X$ coefficients that we use in this work, ignoring all the off-diagonal terms.  
The actual lattice QCD calculations have statistical error bars on their calculations but here we just use the central values (sometimes averaged over collaborations, when possible).

\begin{table}[htbp]
    \centering
    \begin{tabular}{c@{\hspace{20pt}}c@{\hspace{20pt}}c@{\hspace{20pt}}c@{\hspace{20pt}}c}
    \toprule
    $n$ & $\kappa_n^B$ & $\kappa_n^S$ & $\kappa_n^Q$ & $\kappa_n^{BS}$\\
    \colrule
    $2$ & $0.015$  & $0.017$ & $0.029$ & $-0.0050$\\
    $4$ & $0.0007$ & $0.004$ & $0.008$ & n/a \\
    \botrule
    \end{tabular}
    \caption{Values used for the $\kappa^X_n$ coefficients and single available off-diagonal term $\kappa_2^{BS}$ that describe the slope of the pseudo-critical temperature for the diagonal terms in Eq.\ (\ref{eqn:BSQchiral_diag}).  These values are averaged central values from different lattice QCD collaborations \cite{Borsanyi:2011bn,Bellwied:2015rza,Borsanyi:2020fev,HotQCD:2018pds,Ding:2024sux}}
    \label{tab:kappas}
\end{table}

\begin{figure}[htbp]
    \centering
    \includegraphics[width=0.99\columnwidth]{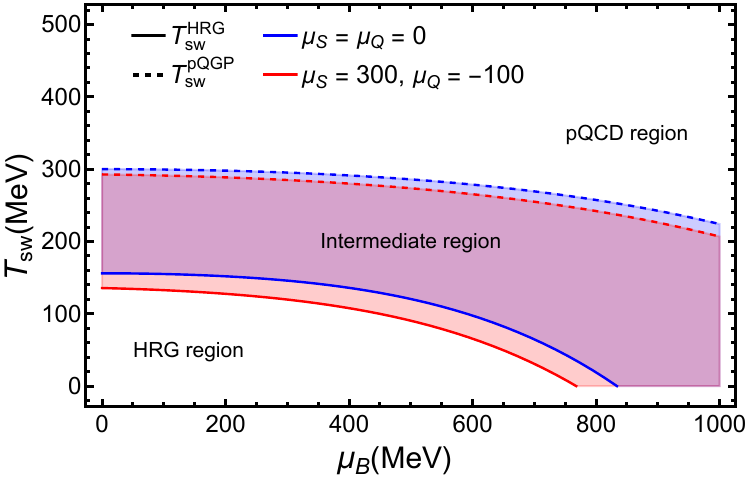}
    \caption{(color online) Intermediate region between HRG and pQGP regimes, where the switching temperatures are defined by \cref{eqn:BSQchiral_diag} and the parameter values in \cref{tab:kappas}, obtained from lattice QCD. In blue, the case of $\mu_S=\mu_Q=0$ and in red, the case of non-vanishing $\mu_S$ and $\mu_Q$. The transition line between the HRG model and the intermediate region is depicted by a solid line and as a dashed line for the transition between intermediate region and the pQCD regime.
    } \label{transitionline}
\end{figure}

\Cref{transitionline} shows the chiral transition line as defined in \cref{eqn:BSQchiral_diag} and the values in \cref{tab:kappas}. 
In this work, we use the form of \cref{eqn:BSQchiral_diag} to mark the boundary between the HRG and the interpolation function that is defined as $T_{\mathrm{sw},0}^\mathrm{HRG} = 156$ MeV and to mark the boundary between the interpolation function and pQCD that is defined as $T_{\mathrm{sw},0}^\mathrm{pQCD} = 300$ MeV. 
We note that the values of  $T_{\mathrm{sw},0}^\mathrm{HRG}$ and $T_{\mathrm{sw},0}^\mathrm{pQCD}$ can easily be varied, but at this point, we are just attempting to put in reasonable guesses for the range of applicability of these models.
However, because the phase transition at vanishing densities is a cross-over, the transition from hadrons into quarks and gluons does not occur at one fixed temperature but rather at a range of temperatures. 
We point out this fact because different transport coefficients can also have inflection points across a range of temperatures; see \cite{Rougemont:2017tlu,Grefa:2022sav} for specific examples.

\subsection{Different temperature regimes and rescaling}
\label{Subsec:Different_temperature_regimes_and_rescaling}

 Temperatures below the switching line ($T_\mathrm{sw}=156$ MeV for $\mu_B=0$ MeV), i.e., $T<T_\mathrm{sw}$, are studied using the hadron resonance gas model with excluded-volume (EV-HRG). 
For the region above the switching line, i.e., $T>T_\mathrm{sw}$, one expects that pQCD would be the most appropriate description, however, it has been shown by \cite{Su:2012iy,Mogliacci:2013mca,Haque:2014rua} that pQCD calculations of thermodynamic observables match lattice calculations at temperatures above $T\gtrsim 300$ MeV for $\mu_B=\mu_S=\mu_Q=0$ MeV, so our lower boundary was set according to these results. 
The intermediate region between $T=156$ MeV to $T=300$ MeV was calculated using an interpolating function. 

As discussed in \cref{pQCD}, shear viscosity for $T>300$ MeV was obtained using kinetic theory as a function of the BSQ chemical potentials and the strong coupling $1/g^4 \log(g^{-1})$. 
As a consequence of the logarithmic dependence of the coupling, it is not possible to reach reasonable values without first performing a rescaling of these results. 
This procedure ensures that our LL calculations match the NLO results for $\eta/s$ obtained by Ghiglieri, Moore, and Teaney (GMT) \cite{Ghiglieri:2018dib}.
These NLO results are the state-of-the-art calculations available but do not include any conserved charge due to the high complexity of the problem.
We begin by extracting the values of $\eta/s$ for NLO for a fixed coupling using the two-loop EQCD (Electrostatic QCD \cite{Braaten:1994na,Braaten:1995cm,Braaten:1995jr,Kajantie:1995dw,Kajantie:1997tt}) value with $\mu_\mathrm{EQCD} = (2.7 \leftrightarrow 4\pi)T$. 
We use these values to match our results at $\mu_B=\mu_S=\mu_Q=0$ MeV and set the scale for each temperature. 
Once the rescaling is done for the pQCD results at vanishing densities, the interpolation function can be calculated by simply matching the pQCD and HRG results.

\subsection{Connecting HRG and pQCD}
\label{Subsec:Connecting_HRG_and_pQCD}

There are infinitely many ways to match the HRG to the rescaled pQCD shear viscosity.  
We generally believe a minimum exists around the cross-over phase transition but have no guidance on the nature of that transition within full QCD with 2+1 quarks. Thus, we choose to use a simple polynomial fit in this work, as defined by
\begin{equation}\label{polynm}
    F(x)=  a+ b x + c x^{2} + d x^{-1}+ f x^{-2},
\end{equation}
where $a$, $b$, $c$, $d$, and $f$ are free parameters that are chosen to ensure that the shear viscosity matches the HRG and pQCD at their respective transition points.
Then, we obtain different types of fits by setting different coefficients to zero. 
For instance, if $c=d=f=0$, we have a linear fit, and using $d=f=0$ leads to a quadratic fit.

We ensure that the transition points must always match, such that, $\eta T/w$ is continuous.
As an example, let us explain the matching for a simple linear fitting:
\begin{equation}\label{eqn:int_linear}
    F(x)=a+bx,
\end{equation}
where we have set $c=d=f=0$ from Eq.\ (\ref{polynm}).
At fixed $\tilde{\mu}$, one takes the transition point from the HRG calculation to the interpolation function at $T_\mathrm{sw}^\mathrm{HRG}(\mu_B)$ where the corresponding shear viscosity is 
\begin{equation}\label{eq:etaTomega_1}
    \left(\frac{\eta T}{w}\right)_1\equiv \frac{\eta T}{w}\left( T_\mathrm{sw}^\mathrm{HRG}(\mu_B),\tilde{\mu}\right),
\end{equation}
and we also define the transition point from the interpolation function to the pQCD shear viscosity at $T_\mathrm{sw}^\mathrm{pQCD}(\mu_B)$ where the corresponding shear viscosity is 
\begin{equation}\label{eq:etaTomega_2}
    \left(\frac{\eta T}{w}\right)_2\equiv \frac{\eta T}{w}\left(T_\mathrm{sw}^\mathrm{pQCD}(\mu_B),\tilde{\mu}\right).
\end{equation}
From \cref{eq:etaTomega_1,eq:etaTomega_2} we can determine $a$ and $b$ by enforcing continuity.  
Since this is a linear function, we know that $b$ is the slope such that we can substitute in our two points at $T_1=T_\mathrm{sw}^\mathrm{HRG}(\mu_B)$ and $T_2=T_\mathrm{sw}^\mathrm{pQCD}(\mu_B)$ to calculate $b$,
\begin{equation}
    b=\frac{\left(\eta T/w\right)_2-\left(\eta T/w\right)_1}{T_{sw,pQCD}-T_{sw,HRG}}.
\end{equation}
We can then substitute in $b$ at either point back into \cref{eqn:int_linear} to obtain $a$,
\begin{equation}
    a=\left(\frac{\eta T}{w}\right)_1-\frac{\left(\eta T/w\right)_2-\left(\eta T/w\right)_1}{T_\mathrm{sw}^\mathrm{pQCD}-T_\mathrm{sw}^\mathrm{HRG}}\; T_\mathrm{sw}^\mathrm{HRG}.
\end{equation}
The same procedure applies for all polynomials of the form of \cref{polynm}. 
We can always determine at least two unknown coefficients by this matching procedure where if there are $N>3$ coefficients, then $N-2$ remain free parameters.

\begin{figure}
    \centering
    \includegraphics[width=0.99\columnwidth]{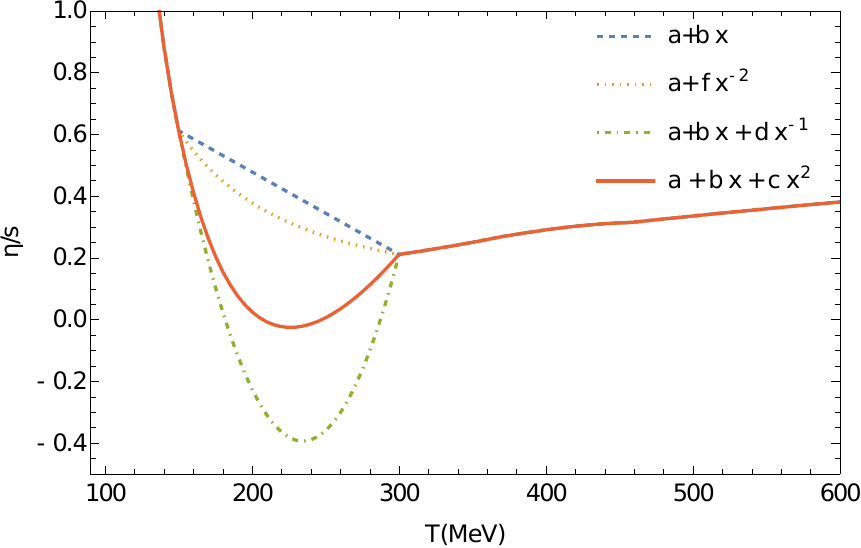}
    \caption{(color online) Different functional forms connecting the HRG shear viscosity at low $T$ and $r=0.25$ fm to rescaled pQCD shear viscosity to match the results from Ref.~\cite{Ghiglieri:2018dib} at high $T$.}
    \label{matching}
\end{figure}

Putting all these pieces together, we can now show different examples of interpolation functions between the two models. 
In \cref{matching}, we compare four different choices of interpolating functions where all have minima in $\eta/s(T)$ at $\tilde{\mu}=0$.
Note that the pQCD results in \cref{matching} are already rescaled according to the GMT results from \cite{Ghiglieri:2018dib}, and we use an excluded-volume HRG with $r=0.25$ fm.  
Comparing the HRG to pQCD at their transition points at $\vec{\mu}=\left\{\mu_B,\mu_S,\mu_Q\right\}=0$, we see that the HRG is somewhat higher but that they are reasonably close to each other such that matching between the two is acceptable.

One can see in \cref{matching} that the location and overall value of the minimum $(\eta/s)_\text{min}$ depends strongly on the functional form of our interpolation function. 
In fact, some functional forms lead to nonphysical values of $\eta/s<0$.  
For the following work, when we obtain $\eta T/w<0$ values, we simply reset $\eta T/w \equiv 0$. 
Thus, we ensure the positivity of shear viscosity, which is consistent with stability constraints for relativistic viscous fluids \cite{Bemfica:2020xym,Almaalol:2022pjc} and statistical mechanics \cite{Callen:1951vq,Kubo:1957mj}. 
However, we generally suggest choosing a functional form at least at $\vec{\mu}=0$ that avoids $\eta/s<0$. 

Out of the different functional forms that we used in \cref{matching}, the form $a+bx+dx^{-1}$ (red curve in \cref{matching}) looks the most like what one would expect from a Bayesian analysis, in that it has a minimum in  $\eta/s$ around the cross-over phase transition. It also avoids any issues with negative values of $\eta/s$.
That being said, that form has an extra free parameter we must select by hand.  
Thus, we choose to take the next by option of $a+bx$ (blue curve in \cref{matching}) that only has two coefficients such that they are both fixed by the transition points. 
Our $a$ and $f$ coefficients are determined independently along slices of $\tilde{\mu}$. 

As a final comment to \cref{matching}, the overall magnitude (especially at high and low $T$) is likely significantly too high compared to what one anticipates from the extraction of $\eta/s(T)$ using Bayesian analyses and the KSS bound from holography \cite{Kovtun:2004de,Bernhard:2019bmu,JETSCAPE:2020mzn,Nijs:2020roc,Parkkila:2021tqq}.
Thus, in the following, we also implement an overall normalization constant $g_\mathrm{norm}$ such that we can shift the functional form up or downward.
One may wonder why we use microscopic approaches and then implement an overall normalization factor.
The reason is that we are primarily interested in extrapolating $\eta T/w$ to finite $\tilde{\mu}$ via models with finite chemical potentials provided $\eta/s$ at $\tilde{\mu}=0$.
Thus, because both the HRG and pQCD shear viscosities have rather non-trivial behaviors, especially considering all three conserved charges, we can explore some interesting behaviors at finite $\tilde{\mu}$.

\begin{table}[htbp]
    \centering
    \begin{tabular}{l@{\hspace{50pt}}l}
    \toprule
    \multicolumn{2}{c}{Free Parameters}  \\
    \colrule
    Parameter           & Default Value\\
    \colrule
    $(\eta/s)_\mathrm{set}$ & 0.08\\
    $T_\mathrm{sw}^\mathrm{HRG}$     & 156 MeV\\
    $T_\mathrm{sw}^\mathrm{pQCD}$    & 300 MeV\\
    $r$              & 0.25 fm\\
    \botrule
    \multicolumn{2}{c}{Constrained Parameters}  \\
    \colrule
    Parameter       & Source \\
    \colrule
    $g_\mathrm{norm}$      & $(\eta/s)_\mathrm{set}/(\eta/s)_\mathrm{min}$ \\
    $g_\mathrm{GMT}$       &  $(\eta/s)_{T}^\mathrm{pQCD}/(\eta/s)_{T}^\mathrm{GMT}$ \cite{Ghiglieri:2018dib}\\
    $a,b,c,d,f$     & determined from matching \\
    \botrule
    \end{tabular}
    \caption{Free and constrained parameters in our phenomenological construction of $\eta T/w$. 
    }
    \label{tab:parameters}
\end{table}

In summary, our shear viscosity algorithm is 
\begin{widetext}

\begin{equation}
 \left(\frac{\eta T}{w} \right)_\mathrm{tot}\left(T,\tilde{\mu}\right)=g_\mathrm{norm}\left\{
 \setlength\extrarowheight{10pt}
    \begin{array}{lr}
      \left(\frac{\eta T}{w} \right)_\mathrm{HRG} &  T \leq T_\mathrm{sw}^\mathrm{HRG}, \\
       \left(\frac{\eta T}{w} \right)_\mathrm{intermediate} &  T_\mathrm{sw}^\mathrm{HRG}<T< T_\mathrm{sw}^\mathrm{pQCD}, \\
     g_\mathrm{GMT} \left(\frac{\eta T}{w} \right)_\mathrm{pQCD} &  T\geq T_\mathrm{sw}^\mathrm{pQCD},
    \end{array}
    \right.
\end{equation}    
\end{widetext}
where $g_\mathrm{GMT}$ is the scaling factor to reproduce the pQCD results from \cite{Ghiglieri:2018dib} at $\tilde{\mu}=0$ and $g_\mathrm{norm}$ is the overall normalization constant that can be constrained at $\tilde{\mu}=0$. 

In Table \ref{tab:parameters}, we summarize all the parameters used in our approach. 
We have two types of parameters: free and constrained.
For the constrained parameters, $g_\mathrm{norm}$ is constrained by $(\eta/s)_\mathrm{set}$ and the details of the rest of the calculation that leads to a certain minimum of $\eta/s$ at $\mu_B=0$ MeV; $g_\mathrm{GMT}$ is determined from the calculations in Ref.\ \cite{Ghiglieri:2018dib}.
Lastly, at least two of the coefficients $a,b,c,d$, and $f$ are determined from our matching procedure. 
The remaining free parameters are not entirely free. The value of $(\eta/s)_\mathrm{set}$ is guided by experimental data, $T_\mathrm{sw}^\mathrm{HRG}$ and $T_\mathrm{sw}^\mathrm{pQCD}$ should be reasonable values that are guided by comparisons of the HRG and pQCD results to lattice QCD, and $r$ is constrained within the given range if one uses PDG2021+ to reproduce lattice QCD thermodynamic results.

\section{Results for \texorpdfstring{$\eta T/w$}{ηT/w}}
\label{Sec:Results_for_etaT_over_omega}

Below, we discuss the results for our phenomenological $\eta T/w$.  
We begin with the results for just one conserved charge, i.e., baryon density, and then consider the implications of all three conserved charges.  
For our results, we cannot show the strangeness neutral case because the interpolation function does not provide information about the densities (since the matching is done at fixed $\tilde{\mu}$) such that the conditions in \cref{eqn:str_neutral} cannot be fulfilled. 
Rather, if one were to use our $\eta T/w$ within a relativistic fluid dynamic code, one would be able to determine the strangeness neutral trajectory via a chosen equation of state.  
In order words, the strangeness neutral trajectory in our approach is equation of state-dependent.

\subsection{\texorpdfstring{$\eta T/w$}{ηT/w} at finite \texorpdfstring{$\mu_B$}{μB} and overall magnitude}
\label{Sec:etaT_over_omega_at_finite_muB_and_overall_magnitude}

 \begin{figure*}[htbp]
\begin{minipage}{.5\textwidth}
    \centering
    \includegraphics[width=0.92\columnwidth]{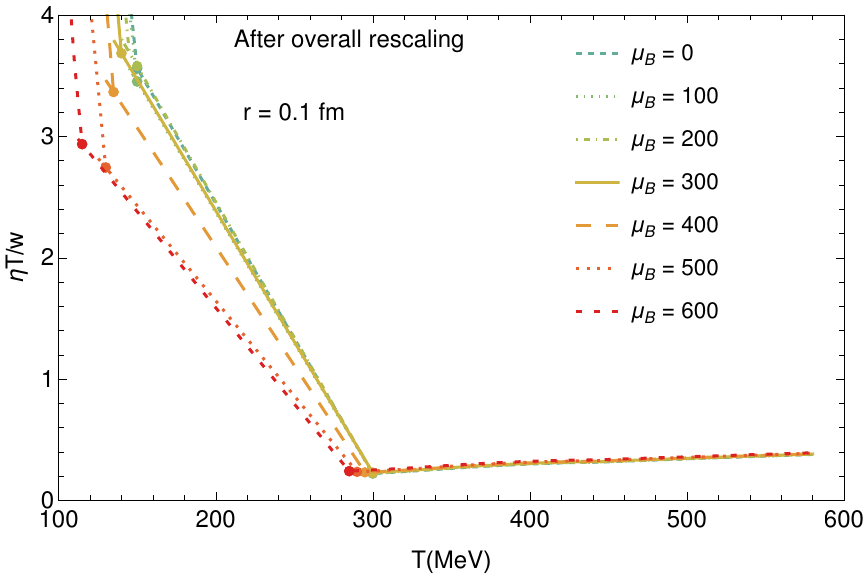}
    \end{minipage}%
    \centering
\begin{minipage}{.5\textwidth}
    \centering
    \includegraphics[width=0.97\columnwidth]{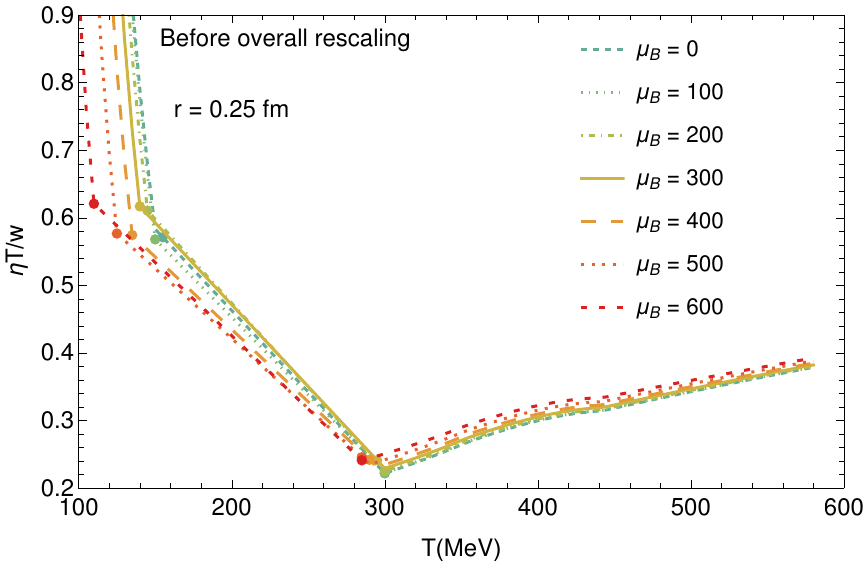}
    \end{minipage}
    \caption{(color online) Left: $\eta T/w$ as a function of the temperature for each value of $\mu_B$ in MeV for $r=0.1$ fm. Right: $\eta T/w$ as a function of the temperature for each value of $\mu_B$ for $r=0.25$ fm.  The results are shown before an overall rescaling is applied (i.e., $g_\mathrm{norm}=1$), but do include the pQCD rescaling. 
    }
    \label{etaT/w_nscaling}
\end{figure*}

 In \cref{etaT/w_nscaling}, we show the result of our $\eta T/w$ for $\mu_B=0$--$600$ MeV where in this case, we have not yet included an overall rescaling, i.e., $g_\mathrm{norm}=1$. 
 The left panel in \cref{etaT/w_nscaling} shows the calculation using  $r=0.1$ fm in the HRG phase, and the right plot shows the $r=0.25$ fm case in the HRG phase.
In both cases, the pQCD sector has the same renormalization to GMT at $\tilde{\mu}=0$ (as described previously).
The region calculated with our interpolation function in  \cref{polynm} is marked with big dots on both edges. 
We find a rather non-trivial difference between $r=0.1$ fm and $r=0.25$ fm in that $r=0.25$ fm leads to a $(\eta T/w)_\mathrm{min}$ at lower temperatures at large $\mu_B$ because its magnitude in the HRG phase is lower. 
That implies that the minimum of $\eta T/w$ shifts to lower values of $T$ as $\mu_B$ increases, which is what we would expect.

Another finding is that for $r=0.1$ fm, all the global minima happen for the pQCD regime at $(\eta T/w)_\mathrm{min}\sim 300$ MeV, which occurs because $\eta T/w$ in the HRG phase is extremely high to the point that it is most likely incompatible with collective flow results calculated using relativistic hydrodynamics. 
Additionally, in the case of $r=0.1$ fm, there is a significant mismatch in the overall magnitude of $\eta T/w$ in the HRG and pQCD limits such that the pQCD effectively has no finite $\vec{\mu}$ dependence when compared to the magnitude of the shear viscosity calculated using the HRG model. 
However, one issue is that the $\eta T/w$ in \cref{etaT/w_nscaling} are still above the KSS bound \cite{Kovtun:2004de} and differ in magnitude from what is seen in Bayesian studies. 
Thus, we will use our scaling factor $g_\mathrm{norm}$ to readjust the magnitude of $\eta/s$ at $\tilde{\mu}=0$ to obtain values in the same order of magnitude as the KSS bound following the procedure in \cite{McLaughlin:2021dph}.
The introduction of this renormalizing factor is motivated by our sources of uncertainty, which include the pQCD NLO calculation, the intermediate region switching parameters, and HRG excluded volume, masses, and widths; the $g_\text{norm}$ factor parametrizes these uncertainties into a single quantity that can be easily modified.
Then, $g_\mathrm{norm}$ remains constant across $\vec{\mu}$.
Our procedure is the following:
\begin{itemize} 
    \item Find the minimum value of $(\eta /s)_\mathrm{min}$ at $\mu_B=0$ MeV. With that, we identify $T_\mathrm{min}$ as being the temperature where $(\eta /s)_\mathrm{min}$ has its minimum value for $\mu_B=0$ MeV.
    \item Then we can determine $g_\mathrm{norm}$ from 
    \begin{equation}
        g_\mathrm{norm}=\frac{(\eta/s)_\mathrm{set}}{(\eta /s)_\mathrm{min}},
    \end{equation}
    where we have chosen $(\eta/s)_\mathrm{set}\equiv 0.08$ to be the new minimum of $\eta /s$ at $\mu_B=0$ MeV in the following results.  
\end{itemize}
Since $(\eta/s)_\mathrm{set}$ is a free parameter in our framework, one could choose other values if needed.
In principle, one could use a data-driven approach to determine $(\eta/s)_\mathrm{set}$ as well as the other free parameters discussed in \cref{tab:parameters}. 
However, we leave such a study for future work.
We note here that the point of this work is not to obtain the perfect behavior of $\eta/s(T)$ at vanishing densities (because of all the caveats with the overall magnitudes of $\eta/s$ coming directly from our models) but rather to use these models to obtain a reasonable behavior across $\mu_B$, $\mu_S$, and $\mu_Q$ since there is very little guidance from both theory and experiments on $\eta T/w$ at finite densities.

\begin{figure}[htbp]
    \centering
    \includegraphics[width=0.99\columnwidth]{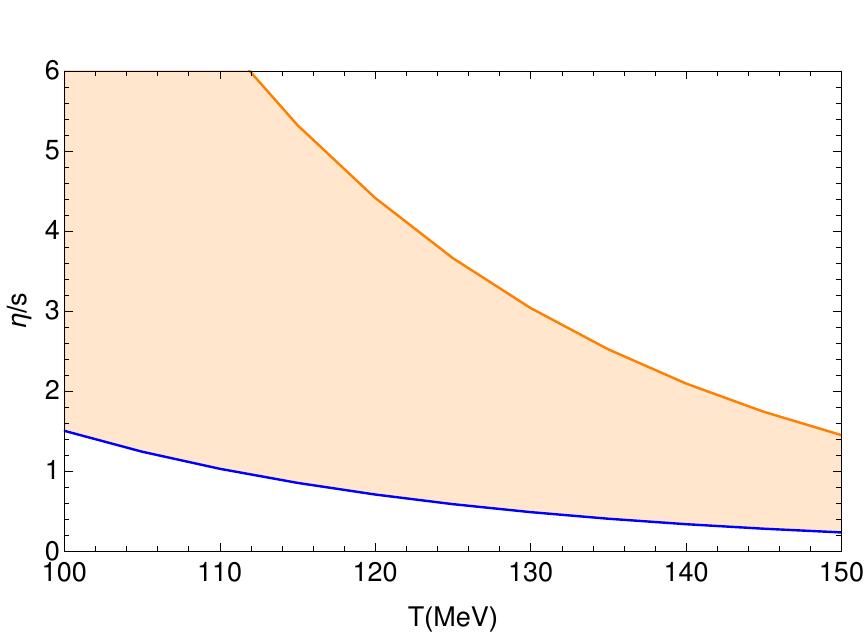}
    \caption{(color online) $\eta/s$ as a function of the temperature for (orange) $r=0.1$ fm and (blue) for $r=0.25$ fm. The shaded region corresponds to the uncertainty in our calculations due to the effective core radius choice.}
    \label{fig:full_etas_norescale}
\end{figure}

From this point on, we will set $(\eta/s)_\mathrm{set}=0.08$ such that $g_\mathrm{norm}\neq 1$.
We emphasize here that the value of $g_\mathrm{norm}$ only shifts the overall magnitude of $\eta T/w$ but does not change the behavior of the curves across the $T$ and $\mu_B$, such that, this behavior comes from the models themselves. 
\Cref{fig:full_etas_norescale} shows the renormalized $\eta/s$ at $\tilde{\mu}=0$ for the two different values of  $r=0.1$ fm (orange) and  $r=0.25$ fm (blue). One can easily see that even after the overall rescaling, the values of $\eta/s$ in the hadronic phase for $r=0.1$ fm are still very large compared to what is expected from other hadronic approaches like SMASH \cite{Hammelmann:2023fqw}. 
Thus, for the remainder of this work, we focus only on the $r=0.25$ fm case, which leads to results that are more reasonable in magnitude and have a relevant finite $\vec{\mu}$ dependence for the pQCD limit.

\begin{figure}[htbp]
    \centering
\includegraphics[width=0.99\columnwidth]{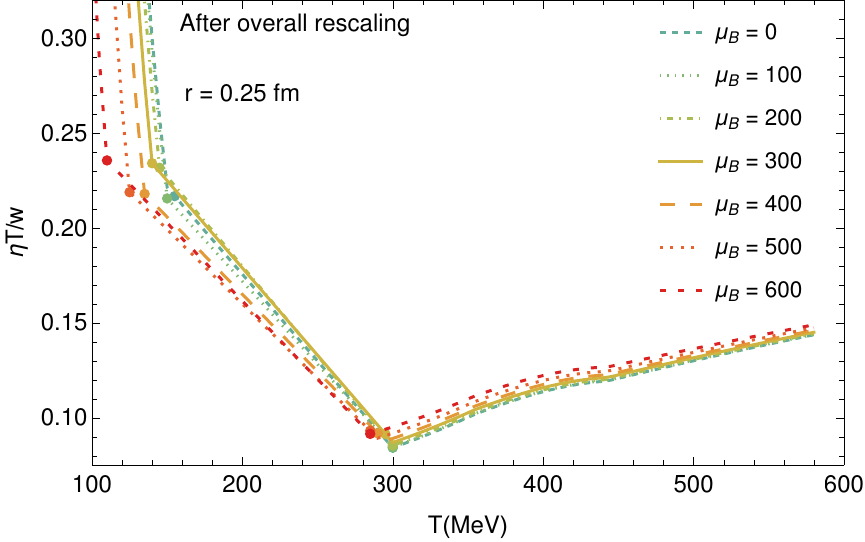}
    \caption{(color online) $\eta T/w(T,\mu_B)$ along different slices of $\mu_B$ in MeV for $r=0.25$ fm. Here, we show the curves with the overall rescaling $g_\mathrm{norm}$. The large dots show the interpolation region on the plot. 
    }
    \label{etaT/w}
\end{figure}

Using our formalism, we now study our shear viscosity at finite $\mu_B$. 
As explained before, we use the same functional form at finite $\mu_B$, but the coefficients in the functional form are $\mu_B$-dependent since they are determined by matching to pQCD and HRG results at both high and low temperatures.
In \cref{etaT/w}, we plot our results at finite $\mu_B$ for $r=0.25$ fm, where we have included the rescaling. 
We find that our approach leads to a non-trivial behavior.  
Because the shear viscosity in the HRG phase at the transition point is generally higher than the shear viscosity at the pQCD transition point, i.e.,
\begin{equation}
    \frac{\eta T}{w}\left(T_\mathrm{sw}^\mathrm{HRG}(\mu_B),\mu_B\right) > \frac{\eta T}{w}\left(T_\mathrm{sw}^\mathrm{pQCD}(\mu_B),\mu_B\right),
\end{equation}
then, the minimum in $\eta T/w$ for a fixed $\mu_B$ ends up being at $T\sim 300$ MeV.  
Of course, if one used a different type of interpolation function, one could adjust the location of the minimum; this just happens to be a consequence of the specific choice made here. 
Another effect we observe is that, due to the drop in $T_\mathrm{sw}$ at finite $\mu_B$, the $\eta T/w(T_\mathrm{sw}^\mathrm{HRG}(\mu_B),\mu_B)$ increases significantly at large $\mu_B$.  
In contrast, the transition point between pQCD and the interpolation does not vary nearly as strongly with $\mu_B$ and therefore remains close to $\eta T/w(T_\mathrm{sw}^\mathrm{pQCD}(\mu_B),\mu_B)\sim (\eta/s)_\mathrm{set}$.

In \crefrange{etaT/w_nscaling}{etaT/w}, we can also study the dependence of the HRG phase on $T$ and $\mu_B$. 
Generally, the HRG has a significantly steeper slope in $T$ than the pQCD phase. 
Additionally, in our setup, we find that the HRG also has a stronger $\mu_B$ dependence than the pQCD phase. 
For a fixed $T$ but increasing $\mu_B$, we find that $\eta T/w$ for the HRG regime consistently decreases, and for a fixed $\mu_B$ but increasing $T$, the $\eta T/w$ for the HRG regime consistently decreases as well. 
Thus, the HRG regime has the complete opposite behavior compared to the pQCD phase, wherein the pQCD regime increases with increasing $T$ and also $\mu_B$.  
The difference in how $\eta T/w$ behaves in the HRG vs. pQCD regime can be understood as follows: in the hadronic phase, we use a geometric cross-section to account for all the scatterings in the system, as the number of particles increases and the scatterings become more frequent, this system tends to equilibrate faster. 

\begin{figure}[htbp]
    \centering
    \includegraphics[width=0.99\columnwidth]{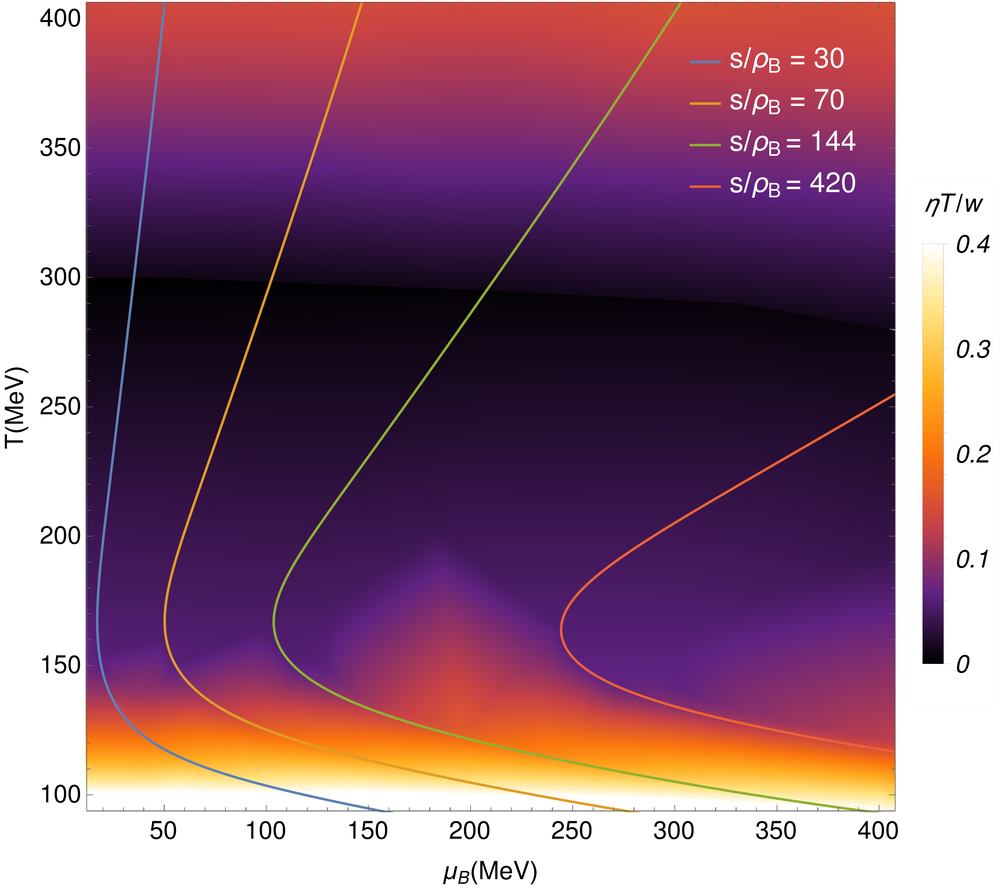}
    \caption{(color online) $\eta T/w$ considering an effective radius $r=0.25$ fm for isentropes trajectories along the QCD phase diagram. The colors indicate the different isentropes defined by $s/\rho_B$. 
    }
    \label{etaT/w_isentropes}
\end{figure}

To summarize our finite $\mu_B$ study, we plot a density plot of $\eta T/w$ across the $(T,\mu_B)$ phase diagram. 
As expected from our other plots, around the phase transition, a minimum in $\eta T/w$ is seen from $T\sim 156$--$300$ MeV.  
At low temperatures, we see a sharp increase in $\eta T/w$ due to the HRG phase.
In the range of $\mu_B=0$--$400$ MeV there is almost no $\mu_B$ dependence but beyond that point ($\mu_B\gtrsim 400$ MeV) we begin to see a stronger $\mu_B$ dependence.

In an ideal fluid, both entropy and baryon number would be conserved, such that, isentropes of entropy density $s$ over baryon density $\rho_B$ would be exactly conserved.
Comparing to isentropes can then provide some insight into how an actual heavy-ion collision would 'experience' viscosity as it expands and cools through the QCD phase diagram. 
The heavy-ion collisions would start at the high $T$, high $\mu_B$ end of the isentrope and as it cools it drops in both $T$ and $\mu_B$ until it hits the phase transition where the isentropes bend back and $\mu_B$ begins to increase again. 
Shortly after the phase transition, the particles are expected to freeze-out, which means the isentrope trajectory is no longer followed down to low $T$ and large $\mu_B$. 
As a quick note, in an actual heavy-ion collision, $s/\rho_B \neq \text{const}.$ due to the entropy production, which is especially relevant around a critical point \cite{Dore:2020jye,Dore:2022qyz,Chattopadhyay:2022sxk}. 
However, even the out-of-equilibrium trajectories are related to the isentropes \cite{Dore:2022qyz}, therefore they can still provide some guidance for what one can expect in experiments.

In \cref{etaT/w_isentropes}, we also plot the isentropes, $s/\rho_B$, for trajectories relevant to the Beam Energy Scan at RHIC. 
The isentropes are shown in colored lines laid on top of the density plot of $\eta T/w$.
Because this range of $\mu_B$ does not yet have a strong $\mu_B$ dependence, then the path from the isentrope is not that different from what is already shown along slices of fixed $\mu_B$. 
Probably the biggest effect from the path of an isentrope vs a slice across $T$ at fixed $\mu_B$ is that for small values of $s/\rho_B$ (corresponding to lower beam energies $\sqrt{s}$) is that the isentrope bends towards large $\mu_B$ as $T$ increases.  
The effect is that the isentropic trajectory remains in the low $\eta T/w$ regime for a longer period of time (in contrast to something that would just increase in $T$).

\subsection{\texorpdfstring{$\eta T/w$}{ηT/w} at finite \texorpdfstring{$\mu_B$}{μB}, \texorpdfstring{$\mu_S$}{μS} and \texorpdfstring{$\mu_Q$}{μQ}}
\label{Subsec:etaT_over_omega_in_the_presence_of_three_conserved_charges}

\begin{figure}[htbp]
    \centering
    \includegraphics[width=0.99\columnwidth]{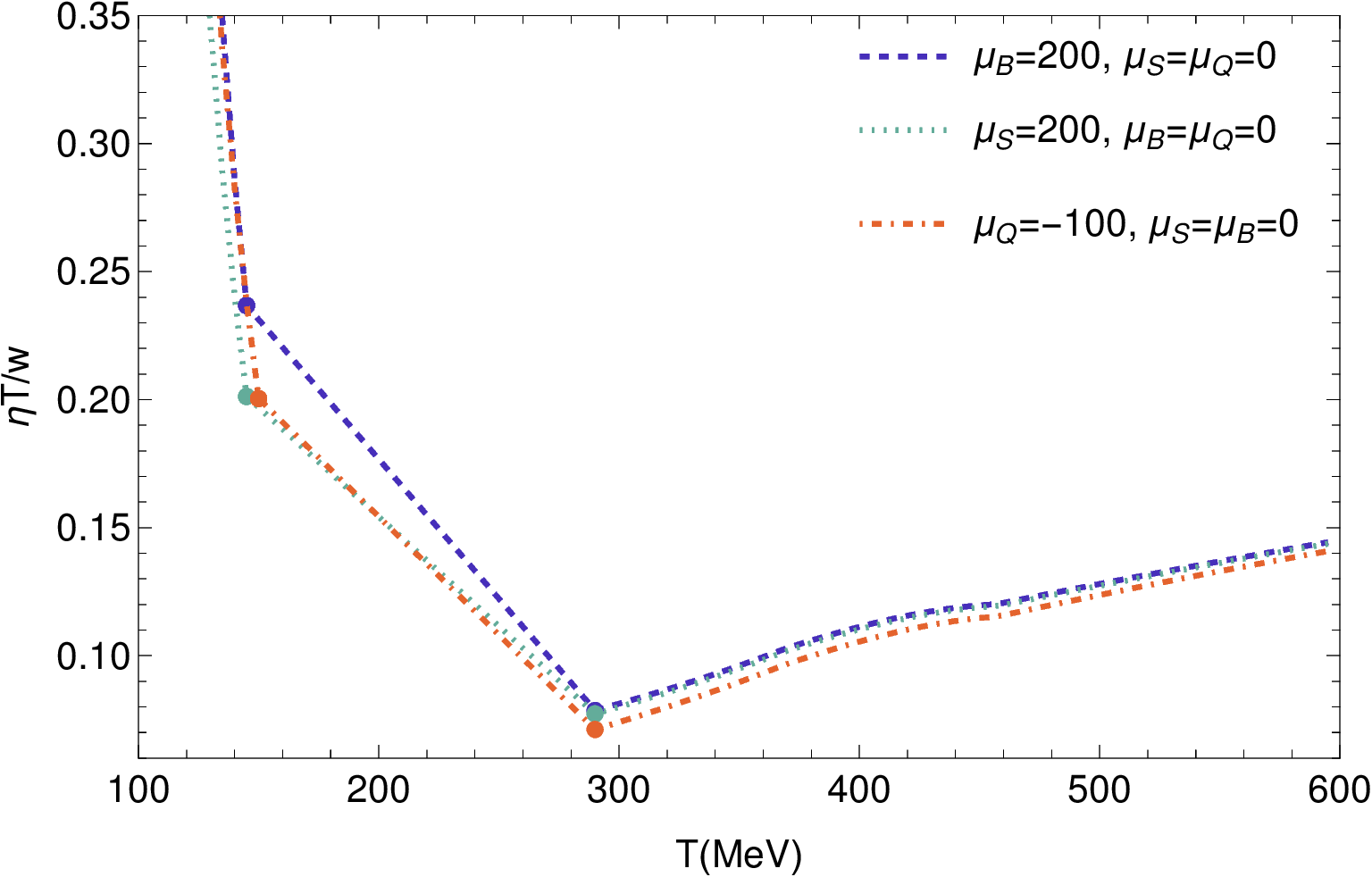}
    \caption{(color online) $\eta T/w$ as a function of the temperature for $r=0.25$ fm comparing the three scenarios of $\mu_B>0$ with $\mu_S=\mu_Q=0$ MeV, $\mu_S>0$ with $\mu_B=\mu_Q=0$ MeV, and $\mu_Q>0$ with $\mu_B=\mu_S=0$ MeV. The large dots show the interpolation region on the plot.
    }
    \label{fig:all_3_finite}
\end{figure}

In this section, we show our final results for shear viscosity along the QCD phase diagram for multiple conserved charges. 
Since there is no good way of displaying a 4D representation of $\eta T/w$, we will build up from one conserved charge up to three. 
We already showed the case where $\mu_B>0$ and $\mu_S=\mu_Q=0$ MeV, shown in \cref{etaT/w,etaT/w_isentropes} in the previous subsection.
We begin in \cref{fig:all_3_finite} by comparing the limits of $\mu_B>0$ with $\mu_S=\mu_Q=0$ MeV, $\mu_S>0$ with $\mu_B=\mu_Q=0$ MeV, and $\mu_Q>0$ with $\mu_B=\mu_S=0$ MeV. 
\Cref{fig:all_3_finite} is similar to what was previously shown for pQCD in \cref{fig:mu_s_shear} but here we have matched to the HRG. 
For the range of chemical potentials considered here, the pQCD results for $\eta T/w$ are nearly the same for $\mu_B$ is finite and for $\mu_S$ is finite.  
However, the HRG results find that a finite $\mu_S$ leads to a suppression of $\eta T/w$ in the HRG phase. 
Thus, at temperatures $T<300$ MeV, the differences between the chemical potentials become more clear, and there is a hierarchy wherein a finite $\mu_B$ has the largest $\eta T/w$, then finite $\mu_Q$, and a finite $\mu_S$ leads to the smallest $\eta T/w$.
This hierarchy differs from what was seen for pQCD ($\mu_S$ and $\mu_Q$ switch).

We note that we are limited to a smaller range in $\mu_Q$ within our approach than for $\mu_B$ and $\mu_S$.
The reason for this limitation is that a singularity appears in the Bose-Einstein distribution at the point where $\mu_Q\to m_\pi$. 
To understand this better, we can look at the Bose-Einstein distribution when the momentum goes to zero for the pions,
\begin{align}
    f_\pi(k=0)&=\sum_{Q=\pm 1}\frac{1}{e^{E_{\pi}(k)/T-Q\mu_Q/T}-1} \nonumber \\
   &= \frac{1}{e^{\left(m_{\pi}-\mu_Q\right)/T}-1}+\frac{1}{e^{\left(m_{\pi}+\mu_Q\right)/T}-1},
\end{align}
where we have substituted in $E_{\pi}=\sqrt{m^2_{\pi}+k^2}$ in the limit of the momentum going to zero $k\rightarrow 0$ such that $E_{\pi}=m_{\pi}$.
Then, at the point that $m_\pi=\pm\mu_Q$, we obtain a singularity. 
To avoid this problem, we always keep $|\mu_Q|<m_{\pi}$ in this work. 
Analogously, one could find a similar point for kaons, i.e.,
\begin{equation}
    f_K(k=0)= \frac{1}{e^{\left(m_{K}-\mu_Q-\mu_S\right)/T}-1}+\frac{1}{e^{\left(m_{K}+\mu_Q+\mu_S\right)/T}-1},
\end{equation}
where we have a singularity at $m_K=\pm\left(\mu_Q+\mu_S\right)$.
Therefore, we always ensure that $|\mu_Q+\mu_S|< 500$ MeV to avoid any issues with the singularity.
Finally, we point out that baryons do not have this same issue because the Fermi-Dirac distribution has a $(+1)$ factor in the denominator instead of a $(-1)$ from the Bose-Einstein distribution such that no singularity appears.

\begin{figure}[htbp]
    \centering
    \includegraphics[width=0.99\columnwidth]{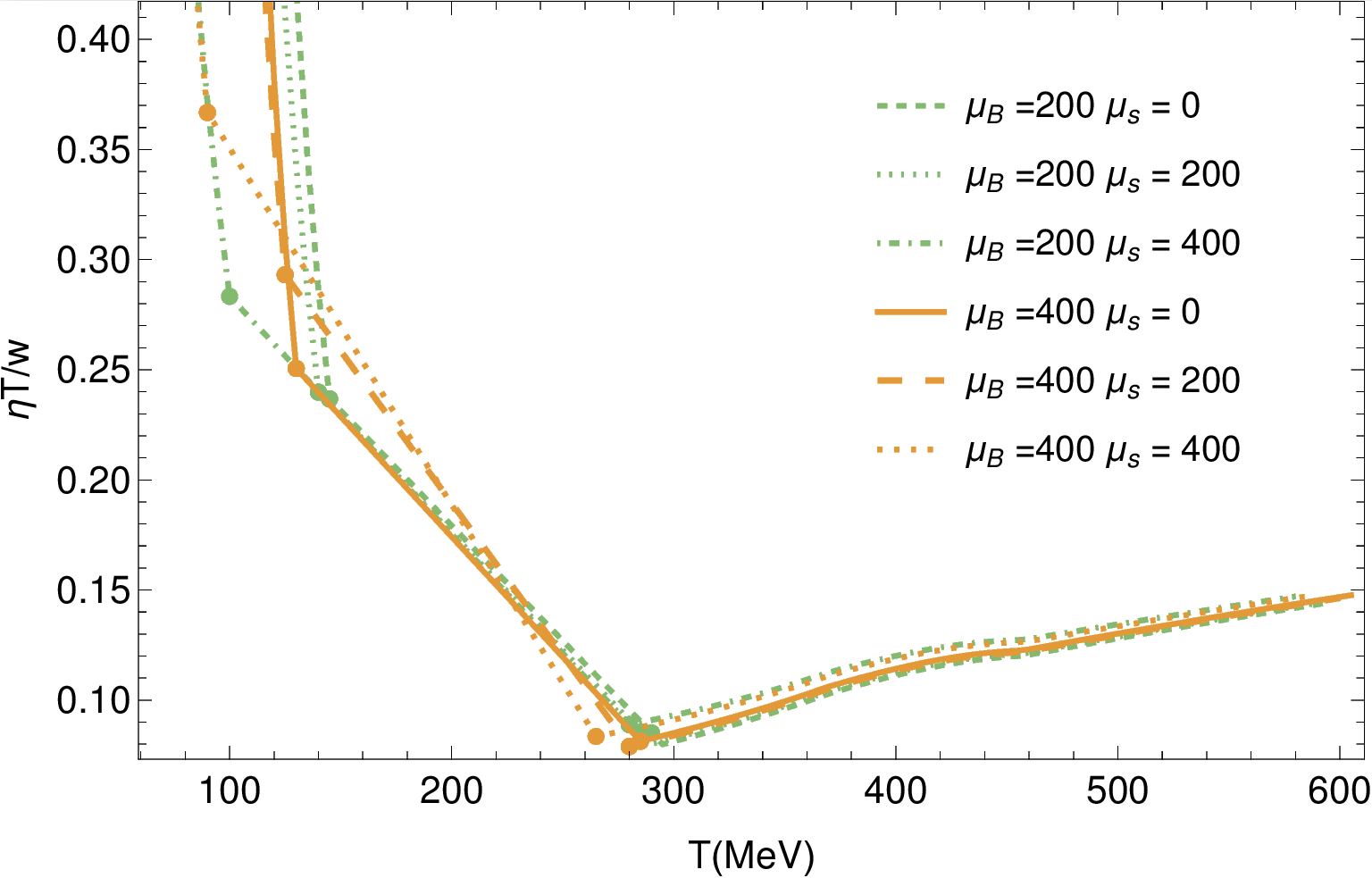}
    \caption{(color online) $\eta T/w$ as a function of the temperature for each value of $\mu_B$ and $\mu_S$ for $r=0.25$ fm. The transition points to the interpolation region are indicated with closed-circle markers.}
    \label{etaT/w_strageness}
\end{figure}

Next, we study a mixture of finite $\mu_B$ and $\mu_S$ in Fig.\ \ref{etaT/w_strageness}.  
There, the green lines are for $\mu_B=200$ MeV, and the orange lines are for $\mu_B=400$ MeV; the variation in $\mu_S$ is shown by different line styles. 
If we first study $\mu_B=200$ MeV, we find that the effect of $\mu_S$ is very small in the range of $\mu_S=0$--$200$ MeV both for the pQCD and HRG regimes, such that there is almost no discernible change. 
However, at $\mu_B=200$ MeV and $\mu_S=400$ MeV, then $\eta T/w$ shifts to higher values in the pQCD regime and lower values in the HRG, effectively flattening out $\eta T/w$ a bit. 
We also see for that combination of $\mu_B$ and $\mu_S$ that there is a strong shift in the transition between interpolation and HRG to lower values of $T$ such that the HRG calculation only appears around $T\sim100$ MeV. 

Now, looking at $\mu_B=400$ MeV, we find that $\mu_S$ plays a larger role when it is switched on.  
The larger $\mu_B$ is much more sensitive to the influence of $\mu_S$. 
The pQCD regime is not so strongly influenced. However, both HRG and pQCD shift to larger $\eta T/w$ when $\mu_S$ is switched on. Furthermore, at both finite $\mu_S$ and $\mu_B$, we see a stronger shift in HRG to lower $T_\mathrm{sw}^\mathrm{HRG}$ throughout the transition from interpolation to HRG.
We can understand this because in the Fermi-Dirac/Bose-Einstein distributions, the influence of $\vec{\mu}$ is normalized by $T$ (i.e., $\vec{\mu}/T$) such that when there is a shift to lower $T$, we expect a stronger influence from changes in $\vec{\mu}$.
We note that the shift in $T_\mathrm{sw}^\mathrm{pQCD}$ is much smaller for a fixed $\vec{\mu}$ than it is for $T_\mathrm{sw}^\mathrm{HRG}$ because the Taylor series in \cref{eqn:BSQchiral_diag} scales with $\mu/T_0$ where the higher value of $T_{\mathrm{sw},0}^\mathrm{pQCD}$ implies that a larger $\mu$ is needed to see the same effect (compared to $T_\mathrm{sw}^\mathrm{HRG}$).
However, the effect of $\mu_B$ in the presence of $\mu_S$ has a stronger influence on the HRG regime.

\begin{figure}[htbp]
    \centering
    \includegraphics[width=0.99\columnwidth]{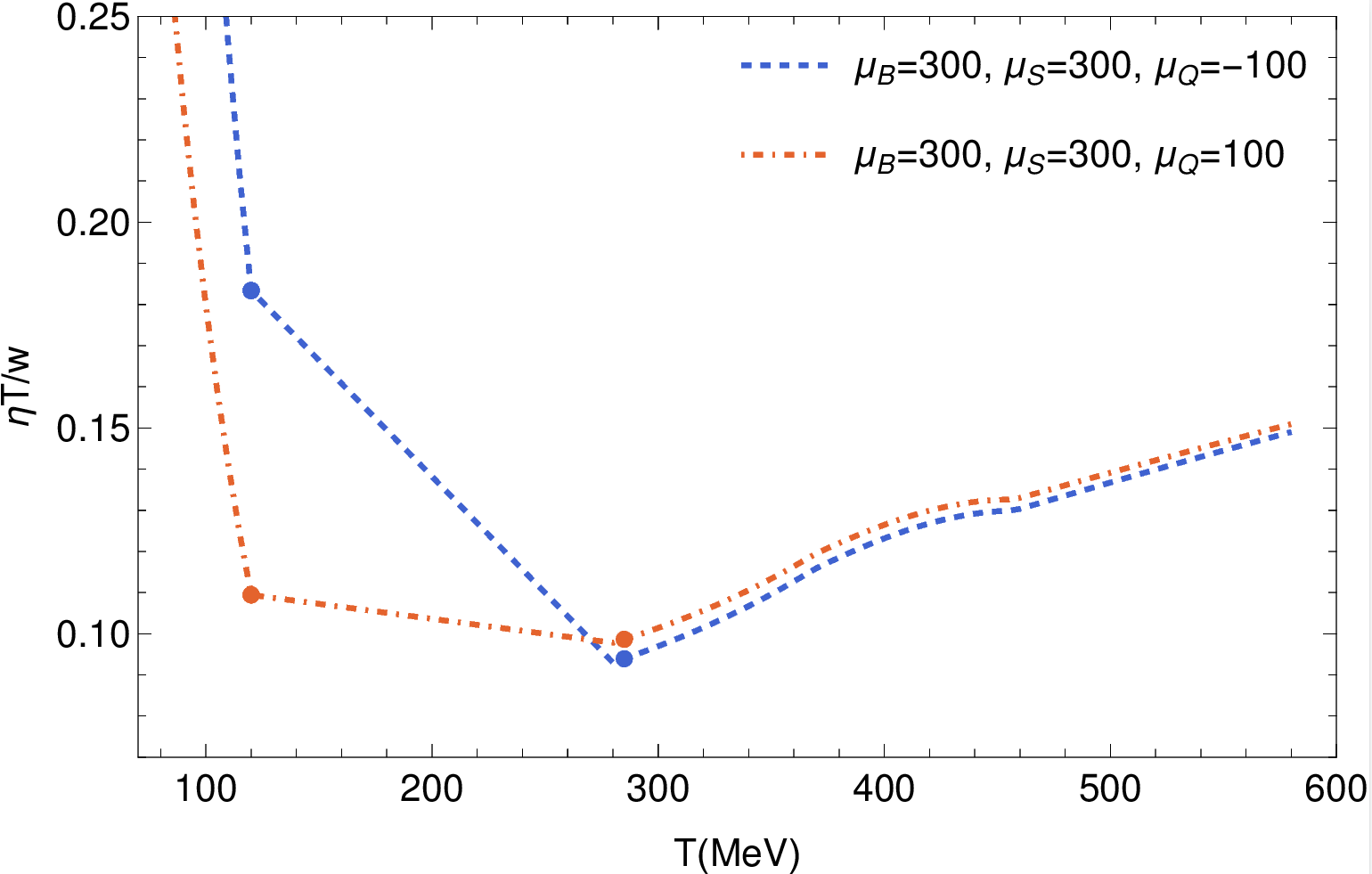}
    \caption{(color online) $\eta T/w$ as a function of the temperature for $r=0.25$ fm when all three chemical potentials are finite. Here we hold $\mu_B=\mu_S=300$ MeV fixed and vary only $\mu_Q=\pm 100$ MeV.
    }
    \label{etaT/w_strageness_muQ}
\end{figure}

Finally, we compare the scenario for $\eta T/w$  when all three chemical potentials are finite, i.e., $\mu_Q>0$, $\mu_B>0$, and $\mu_S>0$ in \cref{etaT/w_strageness_muQ} plotted as a function of temperature $T$.
In principle, we could plot many combinations of $\mu_B,\mu_S$, and $\mu_Q$, but an easy way to demonstrate the effect of $\mu_Q$ is to keep $\mu_B$ and $\mu_S$ fixed and just vary $\mu_Q$.  
Here, we choose positive and negative values of $\mu_Q$ to understand the interplay between chemical potentials. 
Additionally, we are careful to ensure that the combination of $|\mu_S+\mu_Q|<500$ MeV to avoid any issue from the singularity coming from kaons.
We find in \cref{etaT/w_strageness_muQ} that $\mu_Q<0$ has an overall larger $\eta T/w$ than $\mu_Q>0$ in the HRG phase, and the opposite happens in the pQCD phase (the influence is largest in the HRG phase). 
Thus, a positive, large $\mu_Q$ has a flatter $\eta T/w$ across $T$ compared to a negative $\mu_Q$.
Overall, it is clear that when one allows for fluctuations in BSQ conserved charges, it is possible to produce a rather non-trivial dependence on $\vec{\mu}$ for $\eta T/w$. 

\subsection{Fixing \texorpdfstring{$\eta T/w$}{ηT/w} at the transition points}
\label{Sec:Fixing_etaT_over_omega_at_the_transition_points}

Up until this point, we have allowed the calculations of $\eta T/w$ from HRG and pQCD to determine the value of $\eta T/w(T_\mathrm{sw}^\mathrm{HRG}(\mu_B),\mu_B) $ and $ \eta T/w(T_\mathrm{sw}^\mathrm{pQCD}(\mu_B),\mu_B)$ at the transition points. 
Regarding the pQCD results, the overall magnitude is fairly well motivated by NLO calculations. However, convergence has not yet been shown in the series, so NNLO may change the behavior further. 
In the case of the HRG results, the magnitude depends on a correlated combination between the number of hadronic states in the system, their masses, and the excluded-volume extracted from comparisons with lattice QCD.
Additionally, the HRG calculations depend on the assumption of a fixed volume for all hadrons. 
The overall magnitude may change if one were to measure further hadronic states eventually or include more complicated interactions. 
Thus, it is an interesting question how the finite $\vec{\mu}$ behavior looks like if we fix $\eta /s(T_\mathrm{sw}^\mathrm{HRG}(0),0) = \eta /s(T_\mathrm{sw}^\mathrm{pQCD}(0),0)$ at the transition regime at $\vec{\mu}=0$ and allow the finite $\vec{\mu}$ behavior to come from the HRG and pQCD regimes.
In other words, we scale $\eta/s$ from the HRG and pQCD regimes to have the same minimum at $\vec{\mu}=0$ such that we can study how their $\vec{\mu}$ behavior compares to each other.

\begin{figure}[htbp]
    \centering
    \includegraphics[width=0.99\columnwidth]{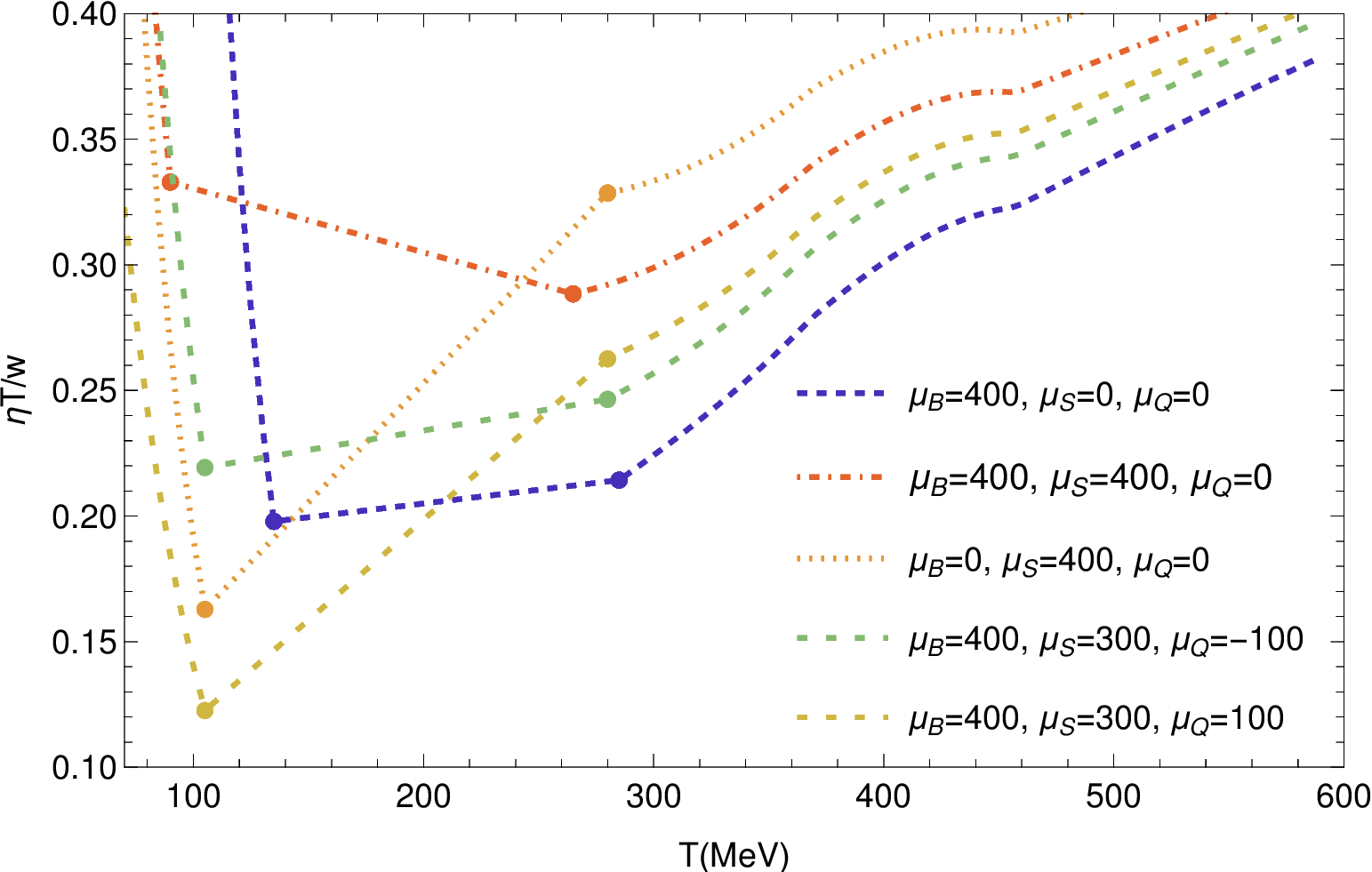}
    \caption{(color online) We scale $\eta/s$ for the HRG and pQCD to be identical at the transition points for $\vec{\mu}=0$, i.e., $\eta /s(T_\mathrm{sw}^\mathrm{HRG}(0),0) = \eta /s(T_\mathrm{sw}^\mathrm{pQCD}(0),0)$. Then, $\eta T/w$ is shown as a function of the temperature in MeV for $r=0.25$ fm for a variety of $\vec{\mu}$.  The large dots show the interpolation region on the plot.
    }
    \label{fig:fixed_transpoint}
\end{figure}

\begin{table}[htbp]
    \centering
    \begin{tabular}{c@{\hspace{12pt}}c@{\hspace{12pt}}c}
    \toprule
    $\mu_B$ (MeV)   & $\mu_S$ (MeV)     & $\mu_Q$ (MeV) \\
    \colrule
    $400$             & $0$                 & $0$ \\
    $400$             & $400$               & $0$ \\
    $0$               & $400$               & $0$ \\
    $400$             & $300$               & $-100$ \\
    $400$             & $300$               & $+100$ \\
    \botrule
    \end{tabular}
    \caption{Combination of chemical potentials used in \cref{fig:fixed_transpoint} when rescaling the pQCD and HRG regimes to be of the same overall values at their minima.}
    \label{tab:chempot_order}
\end{table}

In \cref{fig:fixed_transpoint}, we show the results of fixed $\eta/s$ at the transition point at vanishing densities. 
We choose combinations of chemical potentials to highlight different regimes of the phase diagram, shown in \cref{tab:chempot_order}.
These values ensure that neither the pion nor kaon singularity is reached in the HRG phase.
We find that $\eta T/w$ from pQCD grows more slowly with $T$ and has a small bump around $T\sim 400$ MeV for all combinations of chemical potentials.
In contrast, the HRG phase sharply drops when $T$ increases and always monotonically decreases $\eta T/w$. 

Regardless of the combination of chemical potentials, one can see in \cref{fig:fixed_transpoint} that the magnitude of the change across $\vec{\mu}$ is relatively similar for both the HRG vs the pQCD regime when we rescale their overall magnitudes at the transition region.
For our chosen chemical potentials and $T$ range, pQCD ranges from $\eta T/w\sim [0.2,0.4]$ whereas the HRG ranges from $\eta T/w\sim [0.1,0.4]$.
This effect (that the magnitude of change across $\vec{\mu}$ is similar for pQCD and HRG) was not evident in the previous plots because of our approach algorithm that is motivated both by the physics of pQCD at vanishing densities and also results from Bayesian analyses that demonstrate a nearly flat $\eta/s$ at $\vec{\mu}=0$. 
Thus, our previous results allowed for the pQCD regime to have a much smaller overall magnitude of $\eta T/w$ such that its scaling across $\vec{\mu}$ was suppressed. 
At the time of writing this paper, we do not know what the ``correct" answer is. 
Should the HRG and pQCD regimes have similar $\eta T/w$ values? 
Or does the HRG phase have a larger overall value? 
The answer to that question will then dictate how strongly $\eta T/w$  varies with $\vec{\mu}$ in the high $T$ regime (and also will influence the interpolation regime). 

We can then compare the interplay between the $\mu_B$, $\mu_S$, and $\mu_Q$ behavior between the HRG and pQCD regimes. 
We can use the blue line in \cref{fig:fixed_transpoint} ($\mu_B=400$ MeV) as a baseline wherein only the baryon chemical potential is switched on. 
Comparing this to the limit of $\mu_S=400$ MeV (orange dotted line), we find that the effect of $\mu_S$ increases $\eta T/W$ in the pQCD regime and decreases $\eta T/W$ in the HRG such that the overall $\eta T/W$ at finite $\mu_S$ has a narrow minimum at $T\sim 100$ MeV and then grows steeply with $T$. 
For the case of $\mu_B=\mu_S=400$ MeV, the HRG regime increases $\eta T/w$.
Finally, we compare the scenario where all three chemical potentials are finite, i.e., $\mu_B=400$ MeV, $\mu_S=300$ MeV, and $\mu_Q=\pm100$ MeV (note that we lower $\mu_S$ to avoid the kaon singularity). 
When we have finite $\mu_B,\mu_S$ and vary $\mu_Q$, we find that $\mu_Q<0$ flattens $\eta T/w$ across $T$ because the HRG and pQCD results are similar in value at the transition points. 
However, finite $\mu_B,\mu_S$ and $\mu_Q>0$ we find a much sharper behavior in $\eta T/w$ across $T$ wherein a minimum of $\eta T/w$ occurs at low $T$.

\section{Conclusions and Outlook}
\label{Sec:Conclusions_and_Outlook}

We used pQCD results with three conserved charges, a hadron resonance gas picture, and a state-of-the-art list of resonances to study the QCD shear viscosity at large baryon densities. 
We have applied a phenomenological approach to produce curves of $\eta T/w (T,\mu_B,\mu_S,\mu_Q)$ across the QCD phase diagram, which can used to feed relativistic viscous hydrodynamic codes simulating collisions at energies covered by the RHIC Beam Energy Scan or BSQ fluctuations of conserved charges at the LHC. 
Here, we considered three scenarios, one conserved charge ($\mu_B\neq 0$ with $\mu_S=\mu_Q=0$ MeV), two conserved charges ($\mu_B\neq 0$ and $\mu_S\neq 0$ with $\mu_Q=0$ MeV) and three conserved charges ($\mu_B\neq 0$ and $\mu_S\neq 0$ and $\mu_Q\neq 0$ MeV). 
With that, we have presented the first study of $\eta T/w$ with three conserved charges along the QCD phase diagram and developed an easily reproducible procedure that could rely on other models to calculate $\eta T/w$ as well (e.g., holography, hadron transport, etc.). 
We believe that this could also be performed for different transport coefficients, such as bulk viscosity, which might be more sensitive to critical scaling \cite{PhysRevE.55.403,Moore:2008ws,Monnai:2016kud,Martinez:2019bsn,Dore:2020jye,Dore:2022qyz}. 

We find a non-trivial relationship between $\mu_B$, $\mu_S$, and $\mu_Q$, especially in the pQCD regime since $\mu_S>0$ generally increases $\eta T/w$, $\mu_Q>0$ decreases $\eta T/w$, and $\mu_B>0$ has a non-monotonic behavior. 
The strangeness neutral trajectory generally has $\mu_B>0$, $\mu_S>0$, and $\mu_Q<0$ such that one may expect that $\eta T/w$ decreases with increasing $\mu_B$ in the pQCD regime. 
However, fluctuations in BSQ are anticipated around the strangeness neutral trajectory such that one requires full relativistic viscous hydrodynamic simulations to truly understand the consequences. 
In the HRG regime, large $\vec{\mu}>0$  generally decreases $\eta T/w$ if one looks at a fixed $T$. 
However, $T_\mathrm{sw}^\mathrm{HRG}$ decreases as one increases $|\vec{\mu}|$ such that $\eta T/w$ is large (since $\eta T/w$ grows rapidly as $T$ decreases). 
While $\eta T/w$ generically decreases as one switch on any of the chemical potentials, how much it decreases depends on if one switches $\mu_B$, $\mu_S$, or $\mu_Q$ ($\mu_S$ shows the strongest suppression). 
Thus, the combination of all three chemical potentials and the effect of the transition region (also with three chemical potentials) leads to non-monotonic changes in $\eta T/w$ across the phase diagram. 

There is still space for improvement, as leading-order or even next-leading-order calculations can also be performed for multiple chemical potentials, giving a more detailed description of shear viscosity in the high-temperature regime. 
Additionally, the HRG regime could be further improved by allowing for different types of interactions (here, we always used a constant excluded-volume across all hadrons). 
However, even with these challenges, our framework opens the doorway to performing Bayesian analyses with $\eta T/w$ at finite $\mu_B$, $\mu_S$, and $\mu_Q$.  
One could perform a Bayesian analysis, for instance, varying the overall normalization constant at $\vec{\mu}=0$, the type of interpolation region, and the transition temperatures at $\vec{\mu}=0$. In \cref{fig:fixed_transpoint}, we show the results of fixed $\eta/s$ at the transition point at vanishing densities. 
We choose combinations of chemical potentials to highlight different regimes of the phase diagram, shown in \cref{tab:chempot_order}.
These values are chosen to ensure that neither the pion nor kaon singularity is reached in the HRG phase.
We find that $\eta T/w$ from pQCD grows more slowly with $T$ and has a small bump around $T\sim 400$ MeV for all combinations of chemical potentials.
In contrast, the HRG phase sharply drops when $T$ increases, and it is always a monotonically decreasing $\eta T/w$. 

Regardless of the combination of chemical potentials, one can see in \cref{fig:fixed_transpoint} that the magnitude of the change across $\vec{\mu}$ is relatively similar for both the HRG vs the pQCD regime when we rescale their overall magnitudes at the transition region.
For our chosen chemical potentials and $T$ range, pQCD ranges from $\eta T/w\sim [0.2,0.4]$ whereas the HRG ranges from $\eta T/w\sim [0.1,0.4]$.
This effect (that the magnitude of change across $\vec{\mu}$ is similar for pQCD and HRG) was not evident in the previous plots because of our approach algorithm that is motivated both by the physics of pQCD at vanishing densities and also results from Bayesian analyses that demonstrate a nearly flat $\eta/s$ at $\vec{\mu}=0$. 
Thus, our previous results allowed for the pQCD regime to have a much smaller overall magnitude of $\eta T/w$ such that its scaling across $\vec{\mu}$ was suppressed. 
At the time of writing this paper, we do not know what the ``correct" answer is. 
Should the HRG and pQCD regimes have similar $\eta T/w$ values? 
Or does the HRG phase have a larger overall value? 
The answer to that question will then dictate how strongly $\eta T/w$  varies with $\vec{\mu}$ in the high $T$ regime (and also will influence the interpolation regime). 

We can then compare the interplay between the $\mu_B$, $\mu_S$, and $\mu_Q$ behavior between the HRG and pQCD regimes. 
We can use the blue line in \cref{fig:fixed_transpoint} ($\mu_B=400$ MeV) as a baseline wherein only the baryon chemical potential is switched on. 
Comparing this to the limit of $\mu_S=400$ MeV (orange dotted line), we find that the effect of $\mu_S$ increases $\eta T/W$ in the pQCD regime and decreases $\eta T/W$ in the HRG such that the overall $\eta T/W$ at finite $\mu_S$ has a narrow minimum at $T\sim 100$ MeV and then grows steeply with $T$. 
For the case of $\mu_B=\mu_S=400$ MeV, the HRG regime increases $\eta T/w$.
Finally, we compare the scenario where all three chemical potentials are finite, i.e., $\mu_B=400$ MeV, $\mu_S=300$ MeV, and $\mu_Q=\pm100$ MeV (note that we lower $\mu_S$ to avoid the kaon singularity). 
When we have finite $\mu_B,\mu_S$ and vary $\mu_Q$, we find that $\mu_Q<0$ flattens $\eta T/w$ across $T$ because the HRG and pQCD results are similar in value at the transition points. 
However, finite $\mu_B,\mu_S$ and $\mu_Q>0$ we find a much sharper behavior in $\eta T/w$ across $T$ wherein a minimum of $\eta T/w$ occurs at low $T$.

\section*{Acknowledgments}

The authors would like to thank Guy D. Moore and Jorge Noronha for instructive conversations. I.D. acknowledges the support by the State of Hesse within the Research Cluster ELEMENTS (Project ID 500/10.006), and support by the Deutsche Forschungsgemeinschaft (DFG, German Research Foundation) through the CRC-TR 211 'Strong-interaction matter under extreme conditions'– project number 315477589 – TRR 211.
J.S.S.M. is supported by Consejo Nacional de Humanidades, Ciencias y Tecnolog\'ias (CONAHCYT) under SNI Fellowship I1200/16/2020.
J.N.H. acknowledges support from the US-DOE Nuclear Science Grant No. DE-SC0020633,  DE-SC0023861, and within the framework of the Saturated Glue (SURGE) Topical Theory Collaboration. The work is also supported from the Illinois Campus Cluster, a computing resource that is operated by the Illinois Campus Cluster Program (ICCP) in conjunction with the National Center for Supercomputing Applications (NCSA), and which is supported by funds from the University of Illinois at Urbana Champaign. 
This work was supported in part by the National Science Foundation (NSF) within the framework of the MUSES collaboration, under grant number OAC-2103680.

\bibliography{inspire,NOTinspire}

\end{document}